\documentclass{aa}

\usepackage[varg]{txfonts}

\usepackage[]{hyphenat}
\usepackage[]{microtype}
\usepackage{amssymb}
\usepackage{float}
\usepackage{bm}
\usepackage{epsfig}
\usepackage{soul}
\usepackage{subcaption}
\usepackage{multirow,balance}
\usepackage{amsfonts}
\usepackage{amsmath}
\usepackage{aas_macros}
\usepackage{graphicx}
\usepackage{natbib}
\usepackage{color}
\usepackage{listings} 
\usepackage{balance}
\usepackage{hyperref} 
\usepackage{url}
\usepackage{cases}
\usepackage{tikz}
\usetikzlibrary{decorations.pathreplacing}

\hypersetup{dvips, colorlinks=true, linkcolor=cyan, citecolor=black, filecolor=black, urlcolor=black}

\newcommand{\ri}{\mathrm{i}}
\newcommand{\re}{\mathrm{e}}
\newcommand{\rd}{\mathrm{d}}
\newcommand{\ra}{\mathrm{a}}
\newcommand{\rb}{\mathrm{b}}
\newcommand{\rc}{\mathrm{c}}
\newcommand{\rD}{\mathrm{D}}
\newcommand{\rg}{\mathrm{g}}
\newcommand{\rf}{\mathrm{f}}
\newcommand{\rr}{\mathrm{r}}
\newcommand{\rs}{\mathrm{s}}
\newcommand{\rt}{\mathrm{t}}

\newcommand{\Fs}{F_{\star}}
\newcommand{\Fsa}{F_{\star}^{\ra}}
\newcommand{\Fsb}{F_{\star}^{\rb}}
\newcommand{\Fsc}{F_{\star}^{\rc}}

\newcommand{\bR}{\bm{\mathcal{E}}}
\newcommand{\oP}{\overline{\Phi}}

\newcommand{\oU}{\overline{U}}
\newcommand{\oUr}{\overline{U}_{\rm r}}

\newcommand{\esoft}{\varepsilon_{\rm soft}}

\begin{document}

\setcounter{tocdepth}{3}

\title
{
The secular evolution of discrete quasi-Keplerian systems.\\II. Application to a multi-mass axisymmetric disc\\around a supermassive black hole
}

\titlerunning{The secular evolution of quasi-Keplerian systems. II. Razor-thin discs.}

\author{
J.-B. Fouvry
\inst{\ref{inst1}}\thanks{Hubble Fellow, \email{fouvry@ias.edu}}
\and
C. Pichon
\inst{\ref{inst2},\ref{inst3}}
\and
P.-H. Chavanis
\inst{\ref{inst4}}
}

\institute{
Institute for Advanced Study, Einstein Drive, Princeton, NJ 08540, USA
\label{inst1}
\and
Institut d'Astrophysique de Paris and UPMC, CNRS (UMR 7095), 98 bis Boulevard Arago, 75014, Paris, France\label{inst2}
\and
Korea Institute for Advanced Study (KIAS) 85 Hoegiro, Dongdaemun-gu, Seoul, 02455, Republic of Korea\label{inst3}
\and
Laboratoire de Physique Th\'eorique (IRSAMC), CNRS and UPS, Univ. de Toulouse, F-31062 Toulouse, France\label{inst4}
}

\date{Received \today /
Accepted --}
\abstract{
The drift and diffusion coefficients of the inhomogeneous multi-mass degenerate Landau equation are computed to describe the self-induced resonant relaxation of a discrete self-gravitating quasi-Keplerian razor-thin axisymmetric disc orbiting a massive black hole while relying on Gauss' method. For a disc-like configuration in our Galactic centre, secular diffusion induces an adiabatic distortion of orbits. When considering a disc composed of multiple masses similarly distributed, the population of lighter stars will gain eccentricity, driving it closer to the central black hole provided the distribution function increases with angular momentum. The quenching of the diffusion of a test star in the vicinity of the black hole due to the divergence of the relativistic precessions (the ``Schwarzschild barrier'') is correctly recovered by the kinetic equation. The dual stochastic Langevin formulation yields consistent results and provides a versatile framework in which to incorporate other stochastic processes.
}

\maketitle

\section{Introduction}
\label{sec:introduction}

The dynamical evolution of stellar clusters in the vicinity of galactic centre's supermassive black holes (SMBH) has triggered some interest over the last couple of decades~\citep{MorrisSerabyn1996,Gillessen2009}, amplified by the recent direct detection of gravitational waves through the coalescence of intermediate mass black holes~\citep{LIGOdetect}. Understanding the dynamics of stars in the vicinity of our galaxy's supermassive black hole is now one of the prime goal of the new generation of interferometers such as Gravity~\citep{Gravity}. 
Within the next decade, the community will also build the LISA observatory\footnote{e.g.~\url{http://elisascience.org/whitepaper/}.} to detect gravitational waves from systems of black holes with masses ranging from a few to $10^8$ $M_\odot$~\citep{eLISA}. 
 
SMBHs absorb stars and debris whose orbits reach its loss-cone, where they are either taken directly into the black hole or close enough to interact strongly with it. The continuous loss of stars reshapes the central stellar distribution~\citep{Genzel2000}, also affecting the secular evolution of the SMBH's mass and spin~\citep[e.g.][]{Volonteri2016}.
This dynamical process triggers different observational signatures depending on the mass of the stars, such as binary capture and hyper-velocity star ejection~\citep{Hills1988}, the tidal heating and disruption of stars~\citep{Rees1976}, gravitational waves generation by inspiraling compact remnants. 
These signatures provide indirect evidence of the existence of the central black hole and can be used to test the theory of gravity in the strong field regime~\citep{Merritt2009}.
It is therefore timely to model the wide range of masses involved in nuclear clusters -- from brown dwarfs up to intermediate black holes -- to understand their long-term dynamics near SMBHs, which should eventually allow us to predict e.g. the rate of extreme mass ratio inspirals (EMRI).

 Recently,~\cite{FouvryPichonMagorrian2017} (hereafter paper I) presented the kinetic equation that describes the secular evolution of a large set of particles of various masses orbiting a supermassive black hole (or a protoplanetary debris disc surrounding a star). This so-called set of Balescu-Lenard and Landau kinetic equations was obtained by simply averaging the BBGKY equations over the fast angle that describes motion along the Keplerian ellipses.\footnote{This approach relies on a long tradition of kinetic theory, starting from the seminal papers of~\cite{Landau1936} and~\cite{Chandrasekhar1942,Chandrasekhar1943I,Chandrasekhar1943II}, followed by~\cite{Balescu1960} and~\cite{Lenard1960}, and using the recent developments of~\cite{LucianiPellat1987},~\cite{Heyvaerts2010} and~\cite{Chavanis2012} (see~\cite{Heyvaerts2017} for a review).} It describes self-consistently the long-term evolution of the distribution of multi mass quasi-Keplerian orbits around the central object. It models the diffusion and drift of their actions, induced through their mutual resonant interaction. Hence, this set is the master equation that describes the secular effects of resonant relaxation~\citep{RauchTremaine1996}, and should now be implemented to predict the joint dynamical evolution of the central SMBH and its 
 nuclear cluster.

Following paper~I, the aim of this paper is now to implement this kinetic equation for the Galactic centre's inner stellar cluster.
Specifically, it will quantify the adiabatic distortion of its orbits, the stalled diffusion of test stars near the ``Schwarzschild barrier''~\citep{Merritt2011}, the induced mass segregation in eccentricity and the corresponding quantitative kinematic signatures. As such, it will also provide a first complete implementation of the inhomogeneous multi-mass Landau formalism in an astrophysical context.

The paper is organised as follows.
Section~\ref{sec:RRdisc} presents quasi-Keplerian discs and introduces the degenerate inhomogeneous Landau kinetic equation describing self-consistently these discs' resonant relaxation.
Section~\ref{sec:selfRRdisc} applies this self-consistent diffusion formalism to the self-induced resonant relaxation of a discrete razor-thin quasi-Keplerian disc.
Section~\ref{sec:LangevinBarrier} investigates the stochastic resonant diffusion of individual test stars, in particular the quenching of the diffusion in the neighbourhood of the central BH, the ``Schwarzschild barrier''.
Section~\ref{sec:conclusion} wraps up.
Appendix~\ref{sec:wirewirepotential} details the method used to compute the interaction potential between two Keplerian wires.

\section{Secular evolution of quasi-Keplerian discs}
\label{sec:RRdisc}

This paper focusses on the long-term dynamics of a razor-thin axisymmetric disc of stars surrounding a central supermassive BH. Section~\ref{sec:discgeometry} briefly recalls the appropriate angle-action coordinates that may be used to describe the motion of particles in such a system. Section~\ref{sec:discmodel} presents the disc model that will be considered throughout the paper, while section~\ref{sec:Landaudisc} introduces the degenerate Landau equation. This equation describes self-consistently the long-term evolution of razor-thin discrete quasi-Keplerian discs induced by finite${-N}$ effects (in the limit where collective effects are not accounted for).

\subsection{The disc's geometry}
\label{sec:discgeometry}

Let us assume that the system takes the form of a razor-thin axisymmetric disc, so that the dimension of configuration space is ${ d \!=\! 2 }$. Following the conventions from paper I, let us introduce the angle-action coordinates
\begin{equation}
(\bm{\theta} , \bm{J}) = (\theta^{\rs} , \theta^{\rf} , J^{\rs} , J^{\rf}) = (\bR , \theta^{\rf}) \, ,
\label{AA_disc}
\end{equation}
where ${ \bR \!=\! (\bm{J} , \theta^{\rs}) }$ corresponds to the coordinates of a given Keplerian wire, that are all conserved along the Keplerian motion induced by the central BH. Here, the angles and actions are respectively given by
\begin{align}
& \, \theta^{\rs} = g \;\;\; ; \;\;\; \theta^{\rf} = w \;\;\; ; \;\;\; J^{\rs} = L \;\;\; ; \;\;\; J^{\rf} = I = L + J_{r} \, ,
\label{AA_disc_II}
\end{align}
where ${ \theta^{\rs} \!=\! g }$, being the slow angle, is conserved along the motion and corresponds to the argument of the periapse, while ${ \theta^{\rf} \!=\! w }$ stands for the mean anomaly and is the fast angle that describes the phase of the particle along its Keplerian motion. Finally, equation~\eqref{AA_disc_II} also introduced $L$ and $J_{r}$ as the angular momentum and radial action of a given orbit~\citep{BinneyTremaine2008}. Here, ${ I \!=\! L \!+\! J_{r} }$ is the fast action associated with the orbit, which is adiabatically conserved during the resonant relaxation. For prograde orbits, the mapping ${ (\bm{\theta} , \bm{J}) \!\mapsto\! \bm{x} }$ is given by~\citep{SridharTouma1999,BinneyTremaine2008}
\begin{equation}
\begin{bmatrix}
x
\\
y
\end{bmatrix}
=
\begin{bmatrix}
\cos (g) & - \sin (g)
\\
\sin (g) & \cos (g)
\end{bmatrix}
\cdot
\begin{bmatrix}
a (\cos (\eta) \!-\! e)
\\
a \sqrt{1 \!-\! e^{2}} \sin (\eta)
\end{bmatrix} \, ,
\label{mapping_xy}
\end{equation}
where the semi-major axis $a$, the eccentricity $e$, and the eccentric anomaly $\eta$ are given by
\begin{equation}
a = \frac{I^{2}}{G M_{\bullet}} \;\;\; ; \;\;\; e = \sqrt{1 \!-\! (L / I)^{2}} \;\;\; ; \;\;\; w = \eta \!-\! e \sin (\eta) \, .
\label{definitions_a_e_eta}
\end{equation}
The mapping from equation~\eqref{mapping_xy} also allows us to obtain the mapping to the polar coordinates ${ (\bm{\theta} , \bm{J}) \!\mapsto\! (R , \phi) }$ as
\begin{equation}
R \!=\! a (1 \!-\! e \cos (\eta)) \; ; \; \phi \!=\! g \!+\! \text{Arg} \big[\! \cos (\eta) \!-\! e \,; \sqrt{1 \!-\! e^{2}} \sin (\eta) \!\big] \, .
\label{mapping_R_phi}
\end{equation}

\subsection{The disc model}
\label{sec:discmodel}

\begin{figure}[!htbp]
\begin{center}
\epsfig{file=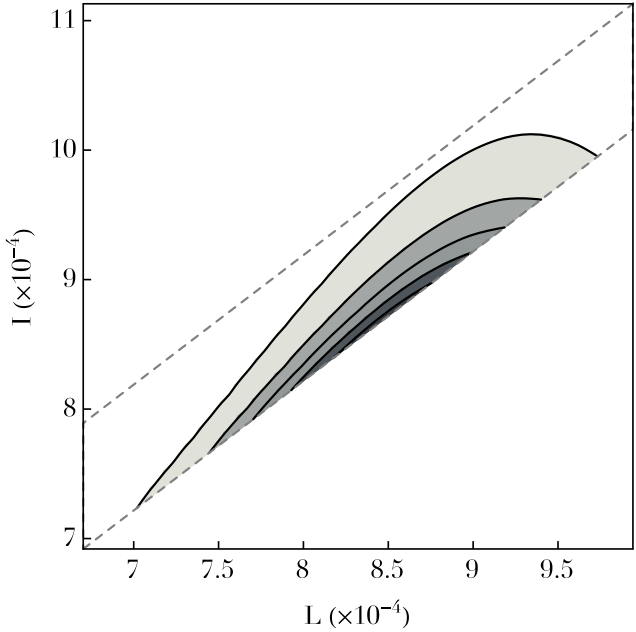,angle=-00,width=0.45\textwidth}
\caption{\small{Illustration of the disc's DF $\Fs$ from equation~\eqref{DF_disc}, in action space ${ \bm{J} \!=\! (L,I) }$. It was assumed here that all stars are prograde, so that ${ L \!>\! 0 }$. Moreover, the action coordinates satisfy ${ I \!\geq\! L }$, so that ${ I \!=\! L }$ corresponds to circular orbits. The contours are spaced linearly between ${ 95\% }$ and ${ 5\% }$ of the DF maximum. The gray dashed lines correspond to the domain in action space from equation~\eqref{definition_action_domain}, to which the computations are restricted.
}}
\label{figDFstar}
\end{center}
\end{figure}

Let us now specify the disc model that will be considered throughout the paper. This disc is chosen to somewhat mimic some of the features of the ``clockwise disc'' of the Galactic centre, considered in~\cite{KocsisTremaine2011}. In order to consider dimensionless quantities, the mass, length and time units are such that
\begin{equation}
M_{\odot} = 1 \;\; ; \;\; 1 \, \text{mpc} = 1 \;\; ; \;\; 1 \, \text{kyr} = 1 \, .
\label{choice_units}
\end{equation}
Within these units, the central BH and the surrounding razor-thin disc are characterised by
\begin{align}
& \, M_{\bullet} = 4 \!\times\! 10^{6} \;\; ; \;\; M_{\star} = 4 \!\times\! 10^{3} \;\; ; \;\; \varepsilon = M_{\star} / M_{\bullet} = 10^{-3} \, ,  \nonumber
\\
& \, \mu_{\star} = 1 \;\; ; \;\; N_{\star} = 4 \!\times\! 10^{3} \, ,
\label{mass_BH_disc}
\end{align}
where $M_{\bullet}$ is the mass of the central BH, $M_{\star}$ the total mass of the disc composed of $N_{\star}$ stars of individual mass $\mu_{\star}$. Because the BH dominates the dynamics, one has ${ \varepsilon \!=\! M_{\star} / M_{\bullet} \!\ll\! 1 }$. For simplicity, let us assume that the star's DF takes the form of a quasi-isothermal DF~\citep{BinneyTremaine2008}, reading
\begin{equation}
\Fs (L ,I) = \frac{1}{M_{\star}} \frac{\Omega_{\rm Kep} (L) \Sigma_{\star} (L)}{\pi \kappa_{\rm Kep} (L) \sigma_{r}^{2} (L)} \, \exp \!\bigg[\! - \frac{\kappa_{\rm Kep} (L)}{\sigma_{r}^{2} (L)} (I \!-\! L) \!\bigg] \, ,
\label{DF_disc}
\end{equation}
which satisfies the normalisation condition ${ \! \int \! \rd \bm{\theta} \rd \bm{J} \, \Fs (\bm{J}) \!=\! 1 }$. Equation~\eqref{DF_disc} introduced the azimuthal and radial orbital frequencies $\Omega_{\rm Kep}$ and $\kappa_{\rm Kep}$~\citep{BinneyTremaine2008}, which in the Keplerian case depend only on $I$ and read
\begin{equation}
\Omega_{\rm Kep} (I) = \kappa_{\rm Kep} (I) = \frac{(G M_{\bullet})^{2}}{I^{3}} \, .
\label{Omega_kappa_Kep}
\end{equation}
In equation~\eqref{DF_disc}, the Keplerian orbital frequencies have to be evaluated in the vicinity of circular orbits, i.e. in ${ I \!=\! L }$. Finally, equation~\eqref{DF_disc} also introduced the local velocity dispersion ${ \sigma_{r} (L) }$ and the disc's surface density ${ \Sigma_{\star} (L) }$. For a Keplerian potential, the mapping between the angular momentum $L$ and the guiding radius $R_{\rg}$ of the corresponding circular orbit is straightforwardly given by
\begin{equation}
R_{\rg} = \frac{L^{2}}{G M_{\bullet}} \, .
\label{link_Rg_L}
\end{equation}
Relying on this mapping, the disc's surface density, $\Sigma_{\star}$, expressed as a function of radius takes the form
\begin{equation}
\Sigma_{\star} (R) = \frac{1}{2 \pi} \frac{M_{\star}}{\sqrt{2 \pi \sigma_{\Sigma}^{2}}} \frac{1}{R} \exp \!\bigg[\! - \frac{(R \!-\! R_{\Sigma})^{2}}{2 \sigma_{\Sigma}^{2}} \!\bigg] \, ,
\label{def_Sigmastar}
\end{equation}
where $R_{\Sigma}$ is the mean radius of the disc, and $\sigma_{\Sigma}$ its radial extent. Such a surface density satisfies very closely the constraint ${ \! \int \! \rd R R \rd \phi \Sigma_{\star} \!=\! M_{\star} }$. Finally, in units of equation~\eqref{choice_units}, we choose
\begin{equation}
R_{\Sigma} = 0.4 \, \text{pc} = 400 \;\;\; ; \;\;\; \sigma_{\Sigma} = 0.15 \!\times\! R_{\Sigma} \, .
\label{choice_RSigma_sigmaSigma}
\end{equation}
In equation~\eqref{DF_disc}, the radial velocity dispersion $\sigma_{r}$ is chosen to be
\begin{equation}
\sigma_{r} = \text{cst.} = 0.2 \!\times\! v_{\rc} (R_{\Sigma}) \, ,
\label{choice_sigma_r}
\end{equation}
where ${ v_{\rc} (R_{\Sigma}) \!=\! L (R_{\Sigma}) / R_{\Sigma} }$ stands for the circular velocity at the radius $R_{\Sigma}$. The larger $\sigma_{r}$, the hotter the disc, and therefore the more eccentric the orbits.
In order not to consider a domain of infinite extent in the ${ (L,I)-}$plane, in all the subsequent numerical applications, we will restrict ourselves to the trapezoidal region
\begin{equation}
L_{\rm min} \leq L \leq L_{\rm max} \;\; ; \;\; L \leq I \leq L \!+\! J_{r}^{\rm max} \, .
\label{definition_action_domain}
\end{equation}
In equation~\eqref{definition_action_domain}, it is bounded in angular momentum by
\begin{equation}
L_{\rm min} = L \big[ R_{\Sigma} \!-\! 2.5 \!\times\! \sigma_{\Sigma} \big] \;\; ; \;\; L_{\rm max} = L \big[ R_{\Sigma} \!+\! 2.5 \!\times\! \sigma_{\Sigma} \big] \, ,
\label{bound_L}
\end{equation}
using the mapping ${ L \!=\! L [R_{\rg}] }$ from equation~\eqref{link_Rg_L}. In addition, in equation~\eqref{definition_action_domain}'s domain, the value of $J_{r}^{\rm max}$ is chosen so that the exponential factor in equation~\eqref{DF_disc} is small enough, so that
\begin{equation}
J_{r}^{\rm max} = 3 \frac{\sigma_{r}^{2} (L_{\Sigma})}{\kappa_{\rm Kep} (L_{\Sigma})} \, ,
\label{choice_Jrmax}
\end{equation}
where ${ L_{\Sigma} \!=\! L [R_{\Sigma}] }$. Finally, let us pave the domain of equation~\eqref{definition_action_domain} with a grid of constant step distance ${ \Delta J }$ defined as
\begin{equation}
\Delta J = \frac{L_{\rm max} \!-\! L_{\rm min}}{n_{\rm Grid}} \, ,
\label{choice_DeltaJ}
\end{equation}
where $n_{\rm Grid}$ is an integer which characterises the density of the considered grid. A fairly sparse grid is used given the computational costs associated with the computation of the wire-wire interaction potential (see Appendix~\ref{sec:wirewirepotential}). Derivatives on the grid are computed by finite differences, so that for example, one has ${ [ \partial f / \partial L] (L,I) \!=\! [ f (L \!+\! \Delta J , I) \!-\! f(L \!-\! \Delta J , I) ] / (2 \Delta J) }$. All the subsequent numerical applications were performed with ${ n_{\rm Grid} \!=\! 30 }$. Figure~\ref{figDFstar} illustrates the disc's DF, $F_{\star}$, from equation~\eqref{DF_disc} on the considered grid.
Finally, as detailed in Appendix~\ref{sec:wirewirepotential}, the gravitational interaction potential is softened so that
\begin{equation}
U (|\bm{x}|) = - \frac{G M_{\bullet}}{\sqrt{\big| \bm{x} \big|^{2} \!+\! \esoft^{2}}} \, .
\label{definition_U_softened}
\end{equation}
In all the upcoming applications, the softening length is given by ${ \esoft \!=\! 10^{-3} \!\times\! R_{\Sigma} }$.

\subsection{The degenerate Landau equation}
\label{sec:Landaudisc}

Because it is made of a finite number of stars, the razor-thin disc presented in section~\ref{sec:discmodel} will undergo a self-induced resonant diffusion on secular timescales. This is the process of resonant relaxation~\citep{RauchTremaine1996}. Paper I recently derived the appropriate master equations to describe such a long-term self-consistent and self-induced evolution. These are the inhomogeneous degenerate Balescu-Lenard and Landau equations. For a razor-thin axisymmetric disc, and in the limit where the contributions from the self-gravitating amplification are neglected, resonant relaxation is governed by the inhomogeneous degenerate Landau equation for razor-thin disc~\citep[paper I]{SridharTouma2016III}, which reads
\begin{align}
\frac{\partial \Fs}{\partial \tau} = & \, \frac{\pi}{N_{\star}} \frac{\partial }{\partial L_{1}} \bigg[ \!\! \int \!\! \rd \bm{J}_{2} \, \delta_{\rD} (\Omega^{\rs} (\bm{J}_{1}) \!-\! \Omega^{\rs} (\bm{J}_{2}))  \nonumber
\\
& \, \times \big| A_{\rm tot} (\bm{J}_{1} , \bm{J}_{2}) \big|^{2} \bigg( \frac{\partial }{\partial L_{1}} \!-\! \frac{\partial }{\partial L_{2}} \bigg) \, \Fs (\bm{J}_{1}) \, \Fs (\bm{J}_{2}) \bigg] \, ,
\label{Landau_Kep_disc}
\end{align}
Equation~\eqref{Landau_Kep_disc} describes the self-induced resonant evolution of the disc's DF as a result of its discreteness. Following the notations from paper I, equation~\eqref{Landau_Kep_disc} introduced the rescaled time $\tau$ defined as ${ \tau \!=\! 2 \pi \varepsilon t }$, with ${\varepsilon \!=\! M_{\star} / M_{\bullet} }$. This equation also involves the disc's total bare susceptibility coefficients ${ A_{\rm tot} (\bm{J}_{1} , \bm{J}_{2}) }$, defined as
\begin{equation}
|A_{\rm tot} (\bm{J}_{1} , \bm{J}_{2}) |^{2} = \sum_{m_{L}} |m_{L}| | A_{m_{L} , m_{L}} (\bm{J}_{1} , \bm{J}_{2}) |^{2} \, ,
\label{definition_Atot}
\end{equation}
where the bare susceptibility coefficients ${ A_{m_{L} , m_{L}} (\bm{J}_{1} , \bm{J}_{2}) }$ are given by the Fourier transform in angle of the wire-wire interaction potential $\oU_{12}$. They read
\begin{align}
A_{m_{L} , m_{L}} (\bm{J}_{1} , \bm{J}_{2}) & \, = \!\! \int \!\! \frac{\rd g_{1}}{2 \pi} \frac{\rd g_{2}}{2 \pi} \, \oU_{12} \, \re^{- \ri m_{L} (g_{1} \!-\! g_{2})}  \nonumber
\\
& \, = \!\! \int \!\! \frac{\rd \Delta g}{2 \pi} \, \oU (\bm{J}_{1} , \bm{J}_{2} , \Delta g) \, \re^{- \ri m_{L} \Delta g} \, ,
\label{definition_A_disc}
\end{align}
since for wires belonging to the same orbital plane, the wire-wire interaction potential $\oU_{12}$ only depends on the phase difference between the two pericentres, ${ \Delta g \!=\! g_{1} \!-\! g_{2} }$. Equation~\eqref{definition_A_disc} introduces the wire-wire interaction potential, $\oU_{12}$, given by
\begin{equation}
\oU (\bm{J}_{1} , \!g_{1} , \!\bm{J}_{2} , \!g_{2}) \!=\! \!\!\! \int \!\! \frac{\rd w_{1}}{2 \pi} \frac{\rd w_{2}}{2 \pi} U \big( | \bm{x}_{1} [ \bR_{1} , w_{1}] \!-\! \bm{x}_{2} [\bR_{2} , w_{2}] | \big) \, .
\label{oU_disc}
\end{equation}
Let us emphasise that the additional symmetry of the interaction potential in equation~\eqref{definition_A_disc} is the very reason why the Landau equation~\eqref{Landau_Kep_disc} for discs can be written without any sum on resonance vectors. The effective calculation of the interaction potential from equation~\eqref{oU_disc} remains a difficult numerical computation, which is the bottleneck of all the upcoming numerical applications. Appendix~\ref{sec:wirewirepotential} details how this potential may efficiently be computed in practice, following Gauss' method~\citep{ToumaTremaine2009}.

Equation~\eqref{Landau_Kep_disc} finally involves a resonance condition on the in-plane precession frequency $\Omega^{\rs}$ of the Keplerian wires. In the present context, the precession frequencies are given by
\begin{equation}
\Omega^{\rs} (\bm{J}) = \Omega^{\rs}_{\rm self} (\bm{J}) + \Omega^{\rs}_{\rm rel} (\bm{J}) \, ,
\label{definition_Omegas}
\end{equation}
where $\Omega^{\rs}_{\rm self}$ stands for the mass precession due to the disc's potential, and $\Omega^{\rs}_{\rm rel}$ for the relativistic precession induced by the central BH. Section~\ref{sec:Landaucalculation} details how these frequencies may be computed. We refer the reader to paper I for a detailed discussion of the physical content of the kinetic equation~\eqref{Landau_Kep_disc}.

In order to emphasise the respective contributions of the diffusion tensor and the friction force by polarisation~\citep{Heyvaerts2017}, one can also rewrite the Landau equation~\eqref{Landau_Kep_disc} by explicitly introducing the disc's drift and diffusion coefficients. It becomes
\begin{equation}
\frac{\partial \Fs}{\partial \tau} = \frac{\partial }{\partial L_{1}} \bigg[ A (\bm{J}_{1}) \, \Fs (\bm{J}_{1}) + D (\bm{J}_{1}) \, \frac{\partial \Fs}{\partial L_{1}} \bigg] \, ,
\label{Landau_Kep_disc_AD}
\end{equation}
where the drift and diffusion coefficients ${ A (\bm{J}_{1}) }$ and ${ D (\bm{J}_{1}) }$ are respectively given by
\begin{align}
& A (\bm{J}_{1}\!) \!=\! - \frac{\pi}{N_{\star}} \!\! \int \!\!\! \rd \bm{J}_{2} \delta_{\rD} (\Omega^{\rs} (\bm{J}_{1}\!) \!-\! \Omega^{\rs} (\bm{J}_{2}\!)) \, | A_{\rm tot} (\bm{J}_{1} , \bm{J}_{2}\!) |^{2} \frac{\partial F_{\star}}{\partial L_{2}} \, ,  \nonumber
\\
& D (\bm{J}_{1}\!) \!=\! \frac{\pi}{N_{\star}} \!\! \int \!\!\! \rd \bm{J}_{2} \delta_{\rD} (\Omega^{\rs} (\bm{J}_{1}\!) \!-\! \Omega^{\rs} (\bm{J}_{2}\!)) \, | A_{\rm tot} (\bm{J}_{1} , \bm{J}_{2}\!) |^{2} F_{\star} (\bm{J}_{2}) \, .
\label{definition_A_D_generic}
\end{align}
In order to stress the conservation of the number of particles during diffusion, the Landau equation~\eqref{Landau_Kep_disc} can finally be written as the divergence of a flux as
\begin{equation}
\frac{\partial \Fs}{\partial \tau} = \frac{\partial }{\partial L_{1}} \bigg[ \mathcal{F}_{L} (\bm{J}_{1}) \bigg] = \frac{\partial }{\partial \bm{J}_{1}} \!\cdot\! \bigg[ \bm{\mathcal{F}}_{\rm tot} (\bm{J}_{1}) \bigg] = \text{div} (\bm{\mathcal{F}}_{\rm tot}) \, ,
\label{Landau_Kep_disc_Div}
\end{equation}
where the flux ${ \mathcal{F}_{L} (\bm{J}_{1}) }$ in the ${L-}$direction and the total flux ${ \bm{\mathcal{F}}_{\rm tot} (\bm{J}_{1}) }$ in the ${ (L ,I)-}$space are respectively defined as
\begin{align}
& \, \mathcal{F}_{L} (\bm{J}_{1}) = A (\bm{J}_{1}) \, \Fs (\bm{J}_{1}) + D (\bm{J}_{1}) \, \frac{\partial \Fs}{\partial L_{1}} \, ,  \nonumber
\\
& \, \bm{\mathcal{F}}_{\rm tot} (\bm{J}_{1}) = \bigg( \mathcal{F}_{L} (\bm{J}_{1}) \, , \, 0 \bigg) \, .
\label{definition_flux_Landau}
\end{align}
Note that the diffusion flux ${ \bm{\mathcal{F}}_{\rm tot} (\bm{J}) }$ is always zero in the ${I-}$direction, which corresponds to the adiabatic conservation of the fast action ${ J^{\rf} \!=\! I }$ during the resonant relaxation. Also note that for an isotropic DF, ${ F_{\star} (\bm{J}) \!=\! F_{\star} (I) }$, the drift coefficients, ${ A (\bm{J}_{1}) }$, from equation~\eqref{definition_A_D_generic} exactly vanish. Finally, recall that the equilibrium states of the self-consistent diffusion equation~\eqref{Landau_Kep_disc} are given by the Boltzmann DFs~\citep{Chavanis2012,SridharTouma2016III}, reading
\begin{equation}
F_{\rm eq} (L , I) = C (I) \, \exp \big[\! - \beta \, H_{\rm eq} (L , I) + \gamma L \big] \, ,
\label{equilibrium_Boltzmann}
\end{equation}
where $\beta$ stands for an inverse temperature, $\gamma $ is the Lagrange multiplier associated with the conservation of the total angular momentum. Here, the energy ${ H_{\rm eq} (L ,I) }$ is given by the primitive
\begin{equation}
H_{\rm eq} (L , I) = \!\! \int \!\! \rd L \, \Omega^{\rs} (L , I) \, .
\label{definition_H_eq_Boltzmann}
\end{equation}
Finally, in equation~\eqref{equilibrium_Boltzmann}, the function ${ C (I) }$ is imposed by the initial conditions. Indeed, because of adiabatic invariance ${ \widetilde{F} (I) \!=\! \! \int \! \rd L \, F (L , I , \tau) }$ is conserved throughout the diffusion, so that ${ C(I) }$ is determined by ${ C (I) \!=\! \widetilde{F} (I) / \! \int \! \rd L \, \re^{- \beta H_{\rm eq} (L , I) + \gamma L} }$. In the high temperature limit, ${ \beta \!\to\! 0 }$, the equilibrium distribution reduces to ${ F_{\rm eq} (L , I) \!=\! C (I) \exp [ \gamma L ] }$~\citep{RauchTremaine1996}.

\section{Self-consistent resonant relaxation}
\label{sec:selfRRdisc}

Having specified the properties of the considered discrete quasi-Keplerian disc and the master equation describing self-consistently its self-induced resonant relaxation, let us now detail how the Landau flux from equation~\eqref{Landau_Kep_disc} may be computed.

\subsection{Computing the Landau flux}
\label{sec:Landaucalculation}

Relying on the fact that in razor-thin discs, the wire-wire interaction potential only depends on the pericentre phase shift ${ \Delta g \!=\! g_{1} \!-\! g_{2} }$, one may perform a harmonic expansion of the form
\begin{equation}
\oU (\bm{J}_{1} , g_{1} , \bm{J}_{2} , g_{2}) = \sum_{k} \oU_{k} (\bm{J}_{1} , \bm{J}_{2}) \, \re^{\ri k \Delta g} \, .
\label{harmonic_U}
\end{equation}
One may then compute this harmonic expansion for each pair ${ (\bm{J}_{1} , \bm{J}_{2}) }$ in the grid from equation~\eqref{definition_action_domain}. In the subsequent numerical applications, the Fourier coefficients are computed by FFT using ${ N_{\rm FFT} \!=\! 2^{7} }$ points. The calculation of the harmonic development of the wire-wire interaction potential in equation~\eqref{harmonic_U} allows then for the computation of two quantities: the self-consistent mass precession frequencies and the bare susceptibility coefficients appearing in the resonance condition from equation~\eqref{Landau_Kep_disc}.

Turning to the total precession frequencies $\Omega^{\rs}$ from equation~\eqref{definition_Omegas}, which originate from both the disc mass precession and the relativistic corrections, the self-consistent mass precession frequencies are given by
\begin{equation}
\Omega^{\rs}_{\rm self} (\bm{J}) = \frac{\partial \oP}{\partial L} \, .
\label{def_Omega_self}
\end{equation}
Equation~\eqref{def_Omega_self} involves the self-consistent potential $\oP$ of the disc, and is given by
\begin{align}
\oP (\bm{J}_{1}) & \, = \!\! \int \!\! \rd \bm{J}_{2} \rd \Delta g \, \Fs (\bm{J}_{2}) \, \oU (\bm{J}_{1} , \bm{J}_{2} , \Delta g)  \nonumber
\\
& \, = 2 \pi \!\! \int \!\! \rd \bm{J}_{2} \, \Fs (\bm{J}_{2}) \, \oU_{0} (\bm{J}_{1} , \bm{J}_{2}) \, ,
\label{definition_Phi_self}
\end{align}
 relying on the harmonic development of the interaction potential from equation~\eqref{harmonic_U}.
The 1PN Schwarzschild relativistic precession frequencies induced by the central BH were obtained in Appendix~A of paper I. They read
\begin{equation}
\Omega^{\rs}_{\rm rel} (\bm{J}) = \frac{\partial \oP_{\rm rel}}{\partial L} \, ,
\label{def_Omega_rel}
\end{equation}
where the relativistic potential $\oP_{\rm rel}$, when correctly normalised, is given by
\begin{equation}
\oP_{\rm rel} (\bm{J}) \!=\! \frac{1}{2 \pi} \frac{M_{\bullet}}{M_{\star}} H_{\rm rel}^{\rm 1PN} (\bm{J}) \text{, with} \; H_{\rm rel}^{\rm 1PN} (\bm{J}) \!=\! - \frac{3 (G M_{\bullet})^{4}}{c^{2}} \frac{1}{L I^{3}} \, .
\label{definition_psi_rel}
\end{equation}
The relativistic precession frequencies can then be explicitly computed and read
\begin{equation}
\Omega^{\rs}_{\rm rel} (\bm{J}) = \frac{1}{2 \pi} \frac{M_{\bullet}}{M_{\star}} \frac{3 (G M_{\bullet})^{4}}{c^{2}} \frac{1}{L^{2} I^{3}} \, .
\label{explicit_Omega_rel}
\end{equation}
Equations~\eqref{def_Omega_self} and~\eqref{explicit_Omega_rel} jointly characterise the precession frequencies that come into play in the resonance condition of the Landau equation~\eqref{Landau_Kep_disc}.

Finally, the harmonic expansion from equation~\eqref{harmonic_U} also allows us to evaluate the disc's total bare susceptibility coefficients from equation~\eqref{definition_Atot}, which become
\begin{equation}
\big| A_{\rm tot} (\bm{J}_{1} , \bm{J}_{2}) \big|^{2} = 2 \sum_{k > 0} k \, \big| \oU_{k} (\bm{J}_{1} , \bm{J}_{2}) \big|^{2} \, ,
\label{calculation_A}
\end{equation}
relying on the fact that $\oU$ being real, one has ${ | \oU_{-k} | \!=\! | \oU_{k} | }$.

\begin{figure}[!htbp]
\begin{center}
\epsfig{file=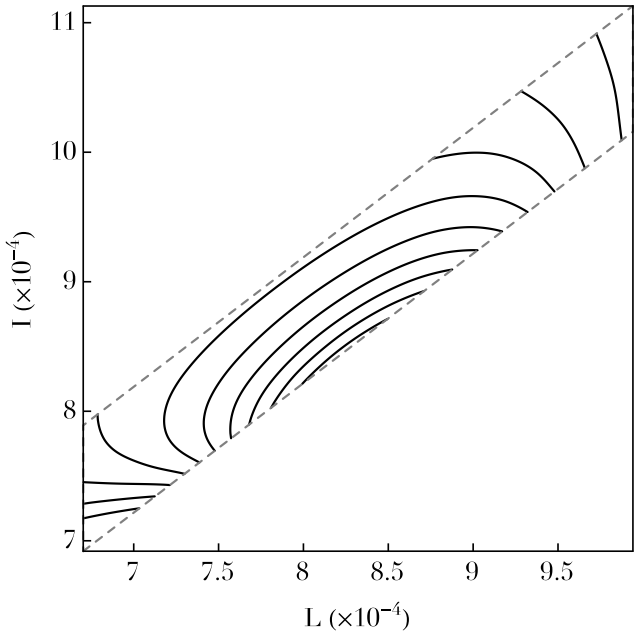,angle=-00,width=0.45\textwidth}
\caption{\small{Illustration of the total precession frequencies ${ \Omega^{\rs} \!=\! \Omega^{\rs}_{\rm self} \!+\! \Omega^{\rs}_{\rm rel} }$ in action space in the neighbourhood of the razor-thin quasi-Keplerian disc introduced in section~\ref{sec:discmodel}. The disc being typically ${ 0.4 \, \text{pc} }$ away from the central BH, the precession frequencies are dominated by the mass precession frequencies $\Omega^{\rs}_{\rm self}$. These mass precession frequencies are retrograde, so that ${ \Omega^{\rs} (\bm{J}) \!<\! 0 }$. The contours in this plot are spaced linearly between ${95\%}$ and ${5\%}$ of the minimum precession frequency satisfying ${ \Omega^{\rs}_{\rm min} \!\simeq\! - 0.2 }$. Because the degenerate Landau equation~\eqref{Landau_Kep_disc} does not involve any resonance vectors, the contours levels of $\Omega^{\rs}$ also correspond to the critical resonant line ${ \gamma (\omega) }$ introduced in equation~\eqref{definition_gamma}.
}}
\label{figContoursOmega}
\end{center}
\end{figure}

Having determined the system's precession frequencies as well as the total bare susceptibility coefficients, the computation of the r.h.s. of equation~\eqref{Landau_Kep_disc} involves dealing with the resonance condition encapsulated by the Dirac delta ${ \delta_{\rD} (\Omega^{\rs} (\bm{J}_{1}) \!-\! \Omega^{\rs} (\bm{J}_{2})) }$, by identifying the critical lines of resonance. To do so, let us first recall the generic definition of the composition of a Dirac delta and a smooth function~\citep{Hormander2003}, which gives
\begin{equation}
\int\limits_{\mathbb{R}^{d}} \!\! \rd \bm{x} \, f (\bm{x}) \, \delta_{\rD} (g (\bm{x})) = \!\! \int\limits_{g^{-1} (0)} \!\!\!\! \rd \sigma (\bm{x}) \, \frac{f (\bm{x})}{|\nabla g (\bm{x})|} \, ,
\label{composition_Dirac}
\end{equation}
where ${ g^{-1} (0) \!=\! \{ \bm{x} \, | \, g (\bm{x}) \!=\! 0 \} }$ is the hypersurface of (generically) dimension ${ (d \!-\! 1) }$ defined by the constraint ${ g (\bm{x}) \!=\! 0 }$, and ${ \rd \sigma (\bm{x}) }$ is its surface measure. In the present context, the resonance condition is given by the function
\begin{equation}
g (\bm{J}_{2}) = \Omega^{\rs} (\bm{J}_{1}) \!-\! \Omega^{\rs} (\bm{J}_{2}) \, .
\label{definition_g_constraint}
\end{equation}
For any given value of $\bm{J}_{1}$, and introducing ${ \omega \!=\! \Omega^{\rs} (\bm{J}_{1}) }$, one may then define the critical resonant curve ${ \gamma (\omega) }$ as
\begin{equation}
\gamma (\omega) = \big\{ \bm{J}_{2} \, \big| \, \Omega^{\rs} (\bm{J}_{2}) \!=\! \omega \big\} \, .
\label{definition_gamma}
\end{equation}
This curve corresponds to the location in action space of all the wires which are in resonance with the precessing wire of action $\bm{J}_{1}$. This is illustrated in figure~\ref{figContoursOmega} for the disc from section~\ref{sec:discmodel}.
Once these resonant lines have been identified, the Landau drift and diffusion coefficients from equation~\eqref{definition_A_D_generic} may straightforwardly be computed, and read
\begin{equation}
A (\bm{J}_{1}) \!=\! \!\! \int_{\gamma (\Omega^{\rs} (\bm{J}_{1}))} \!\!\!\!\!\!\!\!\! \rd \sigma \, \frac{G_{A} (\bm{J}_{1} , \bm{J}_{2})}{| \nabla (\Omega^{\rs} (\bm{J}_{2})) |} \; ; \; D (\bm{J}_{1}) \!=\! \!\! \int_{\gamma (\Omega^{\rs} (\bm{J}_{1}))} \!\!\!\!\!\!\!\!\! \rd \sigma \, \frac{G_{D} (\bm{J}_{1} , \bm{J}_{2})}{| \nabla (\Omega)^{\rs} (\bm{J}_{2}) |} \, .
\label{definition_AD_disc}
\end{equation}
Equation~\eqref{definition_AD_disc} introduced the two functions $G_{A}$ and $G_{D}$ as
\begin{align}
G_{A} (\bm{J}_{1} , \bm{J}_{2}) & \, = - \frac{\pi}{N_{\star}} \big| A_{\rm tot} (\bm{J}_{1} , \bm{J}_{2}) \big|^{2} \, \frac{\partial \Fs}{\partial L_{2}} \, ,  \nonumber
\\
G_{D} (\bm{J}_{1} , \bm{J}_{2}) & \, = \frac{\pi}{N_{\star}} \big| A_{\rm tot} (\bm{J}_{1} , \bm{J}_{2}) \big|^{2} \, \Fs (\bm{J}_{2}) \, ,
\label{definiton_GA_GD}
\end{align}
as well as the resonant contribution ${ |\nabla (\Omega^{\rs} (\bm{J}_{2}))| }$ given by
\begin{equation}
| \nabla (\Omega^{\rs} (\bm{J}_{2})) | = \sqrt{ \bigg[ \frac{\partial \Omega^{\rs}}{\partial L_{2}} \bigg]^{2} \!+\! \bigg[ \frac{\partial \Omega^{\rs}}{\partial I_{2}} \bigg]^{2}} \, .
\label{definition_resonant_contribution}
\end{equation}

\subsection{Self-induced resonant diffusion}
\label{sec:Landausingle}

Equipped with the bricks presented in the previous section, one may then study how the disc's DF, $F_{\star}$, from equation~\eqref{DF_disc} will get to diffuse on secular timescales under the effect of its own discreteness. This involves i) evaluating the pairwise interaction potential $\oU_{12}$ on the grid elements following the Gauss method from Appendix~\ref{sec:wirewirepotential}, ii) determining the precession frequencies (illustrated in figure~\ref{figContoursOmega}), as well as the disc's total bare susceptibility coefficients ${ | A_{\rm tot} |^{2} }$, iii) integrating equation~\eqref{definition_AD_disc} along the associated resonant lines, and iv) computing the disc's self-consistent drift and diffusion coefficients, ${ A (\bm{J}) }$ and ${ D (\bm{J}) }$. These steps allow finally for the computation of the total diffusion flux $\mathcal{F}_{L}$, introduced in equation~\eqref{Landau_Kep_disc_Div}. 

\begin{figure}[!htbp]
\begin{center}
\epsfig{file=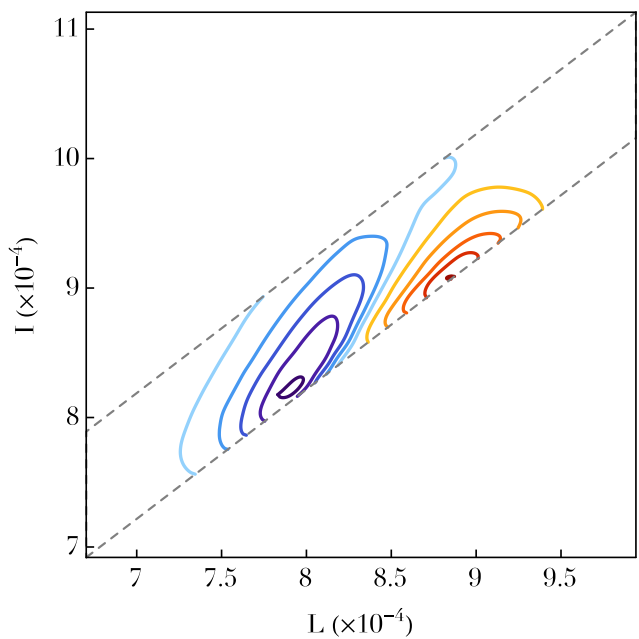,angle=-00,width=0.45\textwidth}
\caption{\small{Illustration of the diffusion flux, $\mathcal{F}_{L}$, predicted by the degenerate Landau equation~\eqref{Landau_Kep_disc_Div} for the razor-thin quasi-Keplerian disc introduced in section~\ref{sec:discmodel}. Following the convention from equation~\eqref{Landau_Kep_disc_Div}, the direction of diffusion of individual particles in action space is given by ${ - \mathcal{F}_{L} }$. Red contours, for which ${ \mathcal{F}_{L} \!<\! 0 }$, correspond to regions where particles tend to diffuse towards larger $L$, i.e. decrease their eccentricity. Blue contours, for which ${ \mathcal{F}_{L} \!>\! 0 }$, are associated with regions in action space, where individual particle tend to diffuse towards smaller $L$, i.e. increase their eccentricity. The contours are spaced linearly between the minimum and the maximum of ${ \mathcal{F}_{L} }$. Within the units of equation~\eqref{choice_units}, the maximum value for the positive blue contours is given by ${ \mathcal{F}_{L}^{\rm max} \!\simeq\! 10^{-10} }$, while the minimum value for the negative red contours reads ${ \mathcal{F}_{L}^{\rm min} \!\simeq\! - 3 \!\times\! 10^{-10} }$.
}}
\label{figFluxSelf}
\end{center}
\end{figure}

The contours of this flux are illustrated in figure~\ref{figFluxSelf}.
Let us first recall that because the equations of motion were averaged w.r.t. the fast Keplerian orbital motion, i.e. w.r.t. $w$ the angle associated with the action $I$, the diffusion is one-dimensional only: individual Keplerian wires conserve their fast action $I$ (i.e. conserve their semi-major axis), and can only diffuse in the ${L-}$direction (i.e. change their eccentricity). In figure~\ref{figFluxSelf}, this translates to the fact that particles diffuse along horizontal lines. Following the convention from equation~\eqref{definition_flux_Landau}, one can note that individual particles will diffuse along the direction of ${ - \mathcal{F}_{L} }$, so that in figure~\ref{figFluxSelf}, most of the individual wires will diffuse towards lower $L$, i.e. towards larger eccentricities. The self-consistent diffusion of the system therefore tends to dynamically heat up the system by making it more eccentric.

\begin{figure}[!htbp]
\begin{center}
\epsfig{file=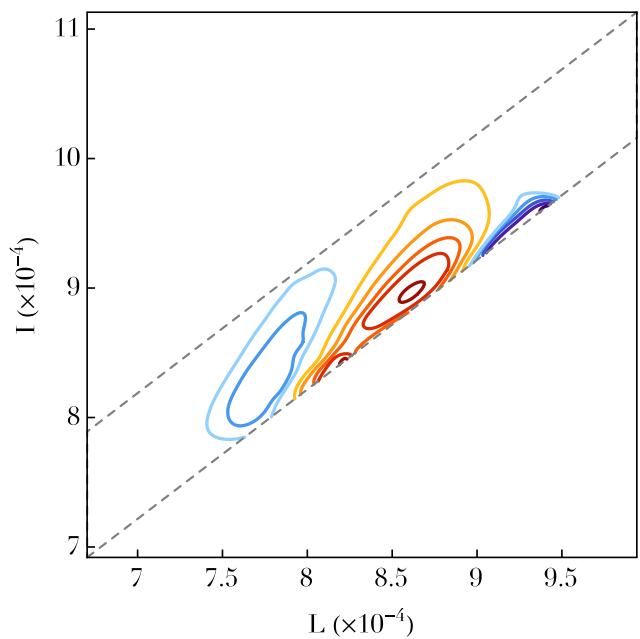,angle=-00,width=0.45\textwidth}
\caption{\small{Illustration of the divergence of the diffusion flux, ${ \text{div} (\bm{\mathcal{F}}_{\rm tot}) }$, predicted by the degenerate Landau equation~\eqref{Landau_Kep_disc_Div} for the razor-thin quasi-Keplerian disc introduced in section~\ref{sec:discmodel}. Red contours, for which ${ \text{div} (\bm{\mathcal{F}}_{\rm tot}) \!<\! 0 }$, correspond to regions from which the wires will be depleted, whereas blue contours, for which ${ \text{div} (\bm{\mathcal{F}}_{\rm tot}) \!>\! 0 }$, are associated with regions in action space, where the value of the disc's DF will increase during the resonant relaxation. The contours are spaced linearly between the minimum and the maximum of ${ \text{div} (\bm{\mathcal{F}}_{\rm tot}) }$. Within the units of equation~\eqref{choice_units}, the maximum value for the positive blue contours is given by ${ \text{div} (\bm{\mathcal{F}}_{\rm tot})_{\rm max} \!\simeq\! 5 \!\times\! 10^{-14} }$, while the minimum value for the negative red contours reads ${ \text{div} (\bm{\mathcal{F}}_{\rm tot})_{\rm min} \!\simeq\! - 10^{-13} }$.
}}
\label{figDivFluxSelf}
\end{center}
\end{figure}

Following the calculation of $\mathcal{F}_{L}$, it is straightforward to compute the divergence of the diffusion flux, ${ \text{div} (\bm{\mathcal{F}}_{\rm tot}) }$, whose contours are illustrated in figure~\ref{figDivFluxSelf}. It is the first application of the degenerate Landau equation in the context of galactic centres, and constitutes a main result of this paper.
This allows us to describe the self-induced local changes of the disc's DF, i.e. to determine the value of ${ [\partial F_{\star} / \partial t ] (t \!=\! 0^{+}) }$. Note from figure~\ref{figDivFluxSelf} that the self-consistent diffusion is associated with an increase in the orbits' eccentricities. It is similar to the localised ``heating'' found in~\cite{FouvryPichonChavanis2015,FouvryPichonMagorrianChavanis2015} when studying the secular self-consistent diffusion of discrete razor-thin self-gravitating stellar discs. There, diffusion induced a heating of the system's DF, which was very localised in action space, taking the form of a narrow resonant ridge. It was amplified by the disc's self-gravitating amplification, as accounted for by the Balescu-Lenard framework. Figure~\ref{figDivFluxSelf} limits itself to the computation of the Landau flux, for which collective effects are not considered. Should the disc be strongly self-gravitating, one expects the self-gravitating amplification not only to accelerate the overall diffusion of the system, but also to enhance it in specific locations in action space where collective effects are the strongest, leading to the appearance of narrow ridges of diffusion.

Let us now estimate the typical timescale associated with this self-consistent resonant diffusion. The contours of $F_{\star}$ presented in figure~\ref{figDFstar} are separated by an increment equal to ${ \Delta F_{\star} \!=\! 0.1 \!\times\! F_{\star}^{\rm max} }$, where ${ F_{\star}^{\rm max} \!\simeq\! 5 \!\times\! 10^{-10} }$ is the maximum of the disc's DF from equation~\eqref{DF_disc}. In order to observe the effects of the secular diffusion, the value of the disc's DF should typically change by an amount of the order of ${ \Delta F_{\star} }$. From the contours of figure~\ref{figDivFluxSelf}, one can note that the maximum of the norm of the divergence of the diffusion flux is typically of the order ${ | \text{div} (\bm{\mathcal{F}}_{\rm tot}) |_{\rm max} \!\simeq\! 10^{-13} }$. Equation~\eqref{Landau_Kep_disc_Div} then allows us to write the relation ${ \Delta F_{\star} \!\simeq\! \Delta \tau_{\rm Ld.} \, | \text{div} (\bm{\mathcal{F}}_{\rm tot}) |_{\rm max} }$, where ${ \Delta \tau_{\rm Ld.} }$ is the typical (rescaled) time during which the Landau equation~\eqref{Landau_Kep_disc_Div} has to be evolved for the disc to undergo a significant diffusion. With the previous numerical values, one gets ${ \Delta \tau_{\rm Ld.} \!\simeq\! 5 \!\times\! 10^{2} }$. Following the convention from equation~\eqref{Landau_Kep_disc}, the associated diffusion time is given by ${ \Delta t_{\rm Ld.} \!=\! \Delta \tau_{\rm Ld.} / ( 2 \pi \varepsilon ) }$, with ${ \varepsilon \!=\! M_{\star} / M_{\bullet} \!=\! 10^{-3} }$. Using the units from equation~\eqref{choice_units}, one finally gets
\begin{equation}
\Delta t_{\rm Ld.} \simeq 100 \, \text{Myr} \, .
\label{value_Delta_t_Landau}
\end{equation}
The self-consistent diffusion captured by the Landau equation~\eqref{Landau_Kep_disc} and computed in figure~\ref{figDivFluxSelf} allows therefore the disc to resonantly diffuse on timescales much shorter than the age of the universe, and also much shorter than the timescale associated with the self-induced relaxation of galactic stellar discs~\citep{FouvryPichonMagorrianChavanis2015}. When accounting for collective effects, the total bare susceptibility coefficients from equation~\eqref{calculation_A} should then be replaced by their dressed analogues. 
As was already observed for non-degenerate stellar discs~\citep{FouvryPichonMagorrianChavanis2015}, provided the disc is sufficiently massive and self-gravitating, one expects that accounting for the wires' polarisation will lead to an acceleration of the disc's self-induced diffusion, and therefore to a reduction of the typical timescale of diffusion from equation~\eqref{value_Delta_t_Landau}.

\begin{figure}[!htbp]
\begin{center}
\epsfig{file=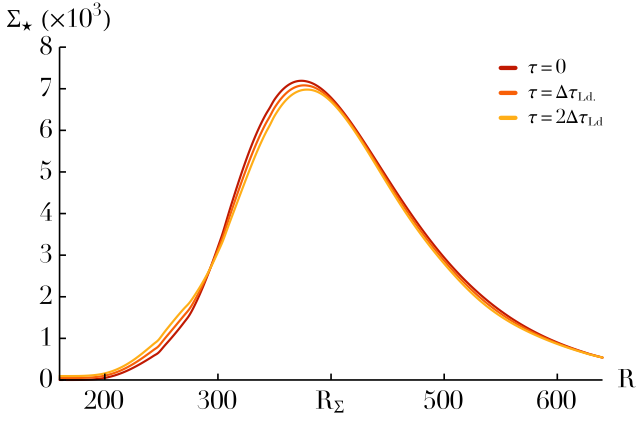,angle=-00,width=0.45\textwidth}
\caption{\small{Illustration of the evolution of the disc's surface density ${ \Sigma_{\star} (R , \tau) }$ as a function of time. As already illustrated in figure~\ref{figDivFluxSelf} in phase space, one can note that on a timescale of the order of ${ \Delta t_{\rm Ld.} }$ (see equation~\eqref{value_Delta_t_Landau}), the disc undergoes a self-induced resonant relaxation which broadens it.
}}
\label{figSigmastarDistorted}
\end{center}
\end{figure}

Following the estimation of ${ \text{div} (\bm{\mathcal{F}}_{\rm tot}) }$ in figure~\ref{figDivFluxSelf}, let us finally investigate how this diffusion impacts the disc's surface density. Recalling the normalisation convention ${ \! \int \! \rd \bm{x} \rd \bm{v} F_{\star} \!=\! 1 }$, the disc's surface density $\Sigma_{\star}$ is given by
\begin{equation}
\Sigma_{\star} (R) = M_{\star} \!\! \int \!\! \rd \bm{v} \, F_{\star} (R , \bm{v}) \, .
\label{calc_init_Sigmastar}
\end{equation}
Appendix~\ref{sec:calcSigma} briefly details how equation~\eqref{calc_init_Sigmastar} may be computed. For sufficiently short diffusion times, the Landau equation~\eqref{Landau_Kep_disc_Div} allows us to approximate the perturbed DF as
\begin{equation}
F_{\star} (\tau) \simeq F_{\star} (\tau \!=\! 0) + \tau \, \text{div} (\bm{\mathcal{F}}_{\rm tot}) \, ,
\label{approx_Landau}
\end{equation}
where the value of the divergence of the diffusion flux is taken for ${ \tau \!=\! 0 }$. One may then use this perturbed DF to estimate the associated perturbed surface density. This is illustrated in figure~\ref{figSigmastarDistorted}, for which the diffusion has been integrated for a time ${ \Delta \tau_{\rm Ld.} }$ as given by equation~\eqref{value_Delta_t_Landau}.
In this figure, one can note that as a result of resonant relaxation, the surface density of the disc gets to diffuse towards smaller radii.

\subsection{Multi-component self-consistent diffusion}
\label{sec:LandauMultidisc}

The previous section considered the self-consistent diffusion of the disc's particles, assuming that all the particles in the disc have the same individual mass. In galactic centres, the range of masses of stars and lighter black holes orbiting the central one is likely to be key to understand the dynamics of the central cluster and possible EMRI.
This section will now show how the Landau equation~\eqref{Landau_Kep_disc}, allows us to describe self-consistently the simultaneous evolution of multiple components. Let us therefore assume that the disc from section~\ref{sec:discmodel} is composed of two distinct components, denoted with the letters ``$\ra$'' and ``$\rb$''. The component ``$\ra$'' is assumed to be composed of $N_{\ra}$ stars of individual mass $\mu_{\ra}$, so that the total mass of this population is given by ${ M_{\star}^{\ra} \!=\! N_{\ra} \mu_{\ra} }$. Similar notations for the component ``$\rb$'' are used. As in section~\ref{sec:discmodel}, the total stellar mass of the system is defined as $M_{\star}$, so that one has ${ M_{\star} \!=\! M_{\star}^{\ra} \!+\! M_{\star}^{\rb} }$. Let us also assume that up to a normalisation the two populations follow the same DF, so that one has ${ \Fsa \!\propto\! \Fsb \!\propto\! \Fs }$, where $\Fs$ stands for the total stellar DF introduced in equation~\eqref{DF_disc}. Keeping track of the normalisations of the multi-component DF (see paper I), the DFs of the components ``$\ra$'' and ``$\rb$'' are then given by
\begin{equation}
\Fsc = \frac{M_{\star}^{\rc}}{M_{\star}} \, \Fs \, ,
\label{relation_DF_multi}
\end{equation}
where the index ``$\rc$'' runs over the two populations ``$\ra$'' and ``$\rb$''. Note that these DFs satisfy the normalisation conventions ${ \! \int \! \rd \bm{x} \rd \bm{v} \Fsc \!=\! M_{\star}^{\rc} / M_{\star} }$. In this multi-component context, the Landau equation~\eqref{Landau_Kep_disc} for razor-thin quasi-Keplerian discs now describes the evolution of each component, and reads
\begin{align}
\frac{\partial \Fsa}{\partial \tau} = & \, \pi \frac{\partial }{\partial L_{1}} \bigg[ \!\! \int \!\! \rd \bm{J}_{2} \, \delta_{\rD} (\Omega^{\rs} (\bm{J}_{1}) \!-\! \Omega^{\rs} (\bm{J}_{2}))  \nonumber
\\
& \!\!\!\!\! \times \big| A_{\rm tot} (\bm{J}_{1} , \bm{J}_{2}) \big|^{2} \, \! \sum_{\rc} \! \bigg( \eta_{\rc} \frac{\partial }{\partial L_{1}} \!-\! \eta_{\ra} \frac{\partial }{\partial L_{2}} \bigg) \, \Fsa (\bm{J}_{1}) \, \Fsc (\bm{J}_{2}) \bigg] \, ,
\label{Landau_Kep_multi}
\end{align}
where the rescaled time $\tau$ is still defined as ${ \tau \!=\! 2 \pi \varepsilon t }$, with ${ \varepsilon \!=\! M_{\star} / M_{\bullet} }$. Equation~\eqref{Landau_Kep_multi} also introduced the small parameter ${ \eta_{\ra} \!=\! \mu_{\ra} / M_{\star} }$, which replaces the factor ${ 1 / N_{\star} }$ present in equation~\eqref{Landau_Kep_disc}. Following equation~\eqref{equilibrium_Boltzmann}, it is straightforward to obtain that the equilibrium of the coupled evolution equations~\eqref{Landau_Kep_multi} are given by the Boltzmann DF reading
\begin{equation}
F_{\rm eq}^{\ra} (L , I) = C^{\ra} (I) \, \exp \big[\! - \beta \, \eta_{\ra} \, H_{\rm eq} (L , I) + \eta_{\ra} \gamma L \big] \, ,
\label{equilibrium_Boltzmann_multi}
\end{equation}
where ${ C^{\ra} (I) }$ are functions imposed by the initial conditions, the inverse temperature $\beta$ and the multiplier $\gamma$ are the same for all the populations, and the energy ${ H_{\rm eq} (L , I) }$ was introduced in equation~\eqref{definition_H_eq_Boltzmann}.

Following equation~\eqref{Landau_Kep_disc_AD}, one can introduce multi-component drift and diffusion coefficients to rewrite equation~\eqref{Landau_Kep_multi} as
\begin{equation}
\frac{\partial \Fsa}{\partial \tau} = \frac{\partial }{\partial L_{1}} \bigg[ \sum_{\rc} \bigg\{ \eta_{\ra} A^{\rc} (\bm{J}_{1}) \, \Fsa (\bm{J}_{1}) + \eta_{\rc} D^{\rc} (\bm{J}_{1}) \, \frac{\partial \Fsa}{\partial L_{1}} \bigg\} \bigg] \, ,
\label{Landau_Kep_disc_multi_AD}
\end{equation}
where the drift and diffusion coefficients ${ A^{\rc} (\bm{J}_{1}) }$ and ${ D^{\rc} (\bm{J}_{1}) }$ depend on the component ``$\rc$'' used as the underlying DF to estimate them. Accounting for normalisations, they read
\begin{align}
A^{\rc} (\bm{J}_{1}) & \!=\! - \pi \!\!\! \int \!\!\! \rd \bm{J}_{2} \, \delta_{\rD} (\Omega^{\rs} (\bm{J}_{1}) \!-\! \Omega^{\rs} (\bm{J}_{2})) \, | A_{\rm tot} (\bm{J}_{1} , \bm{J}_{2}) |^{2} \, \frac{\partial \Fsc}{\partial L_{2}} \, ,  \nonumber
\\
D^{\rc} (\bm{J}_{1}) & \!=\! \pi \!\!\! \int \!\!\! \rd \bm{J}_{2} \, \delta_{\rD} (\Omega^{\rs} (\bm{J}_{1}) \!-\! \Omega^{\rs} (\bm{J}_{2})) \, | A_{\rm tot} (\bm{J}_{1} , \bm{J}_{2}) |^{2} \, \Fsc (\bm{J}_{2}) \, .
\label{AD_multi}
\end{align}
Equation~\eqref{Landau_Kep_disc_multi_AD} can finally be rewritten as
\begin{equation}
\frac{\partial \Fsa}{\partial \tau} = \frac{\partial }{\partial L_{1}} \bigg[ \eta_{\ra} A^{\rm tot} (\bm{J}_{1}) \, \Fsa (\bm{J}_{1}) + D^{\rm tot} (\bm{J}_{1}) \, \frac{\partial \Fsa}{\partial L_{1}} \bigg] \, ,
\label{Landau_Kep_disc_multi_tot_AD}
\end{equation}
where the total drift and diffusion coefficients are
\begin{equation}
A^{\rm tot} (\bm{J}_{1}) = \sum_{\rc} A^{\rc} (\bm{J}_{1}) \;\;\; ; \;\;\; D^{\rm tot} (\bm{J}_{1}) = \sum_{\rc} \eta_{\rc} D^{\rc} (\bm{J}_{1}) \, .
\label{Atot_Dtot_multi}
\end{equation}
Equation~\eqref{relation_DF_multi} assumes that the two populations ``$\ra$'' and ``$\rb$'' follow a DF proportional to the one introduced in equation~\eqref{DF_disc} for the one-component problem. As a consequence, the calculations of the multi-component drift and diffusion coefficients from equation~\eqref{AD_multi} are, up to changes in normalisations, the same as the ones performed in section~\ref{sec:Landaudisc} for the one-component problem. Following the normalisations from equation~\eqref{relation_DF_multi}, the multi-component drift and diffusion coefficients from equation~\eqref{AD_multi} are given by
\begin{equation}
A^{\rc} = \frac{M_{\star}^{\rc}}{M_{\star}} \frac{M_{\star}}{\mu_{\star}} A \;\;\; ; \;\;\; D^{\rc} = \frac{M_{\star}^{\rc}}{M_{\star}} \frac{M_{\star}}{\mu_{\star}} D \, ,
\label{A_D_multi_relations}
\end{equation}
where $A$ and $D$ stand for the drift and diffusion coefficients introduced in equation~\eqref{Landau_Kep_disc_AD} for the one-component problem, and $\mu_{\star}$ is the individual stellar mass of the one-component problem. Following equation~\eqref{Atot_Dtot_multi}, the total drift and diffusion coefficients are then given by
\begin{align}
A^{\rm tot} & \!=\! \bigg[ \frac{M_{\star}^{\ra}}{M_{\star}} \frac{M_{\star}}{\mu_{\star}} \!+\! \frac{M_{\star}^{\rb}}{M_{\star}} \frac{M_{\star}}{\mu_{\star}} \bigg] \!=\! \frac{M_{\star}}{\mu_{\star}} A \, ,  \nonumber
\\
D^{\rm tot} & \!=\! \bigg[ \frac{\mu_{\ra}}{M_{\star}} \frac{M_{\star}^{\ra}}{M_{\star}} \frac{M_{\star}}{\mu_{\star}} \!+\! \frac{\mu_{\rb}}{M_{\star}} \frac{M_{\star}^{\rb}}{M_{\star}} \frac{M_{\star}}{\mu_{\star}} \bigg] D \!=\! \frac{M_{\star}^{\ra} \mu_{\ra} \!+\! M_{\star}^{\rb} \mu_{\rb}}{M_{\star} \mu_{\star}} D \, .
\label{calculation_Atot_Dtot}
\end{align}
These total multi-component drift and diffusion coefficients allow us then to compute the flux appearing in equation~\eqref{Landau_Kep_disc_multi_tot_AD}, given the specific normalisation of the multi-component DFs in equation~\eqref{relation_DF_multi}.

\begin{figure*}[!htbp]
\begin{center}
\epsfig{file=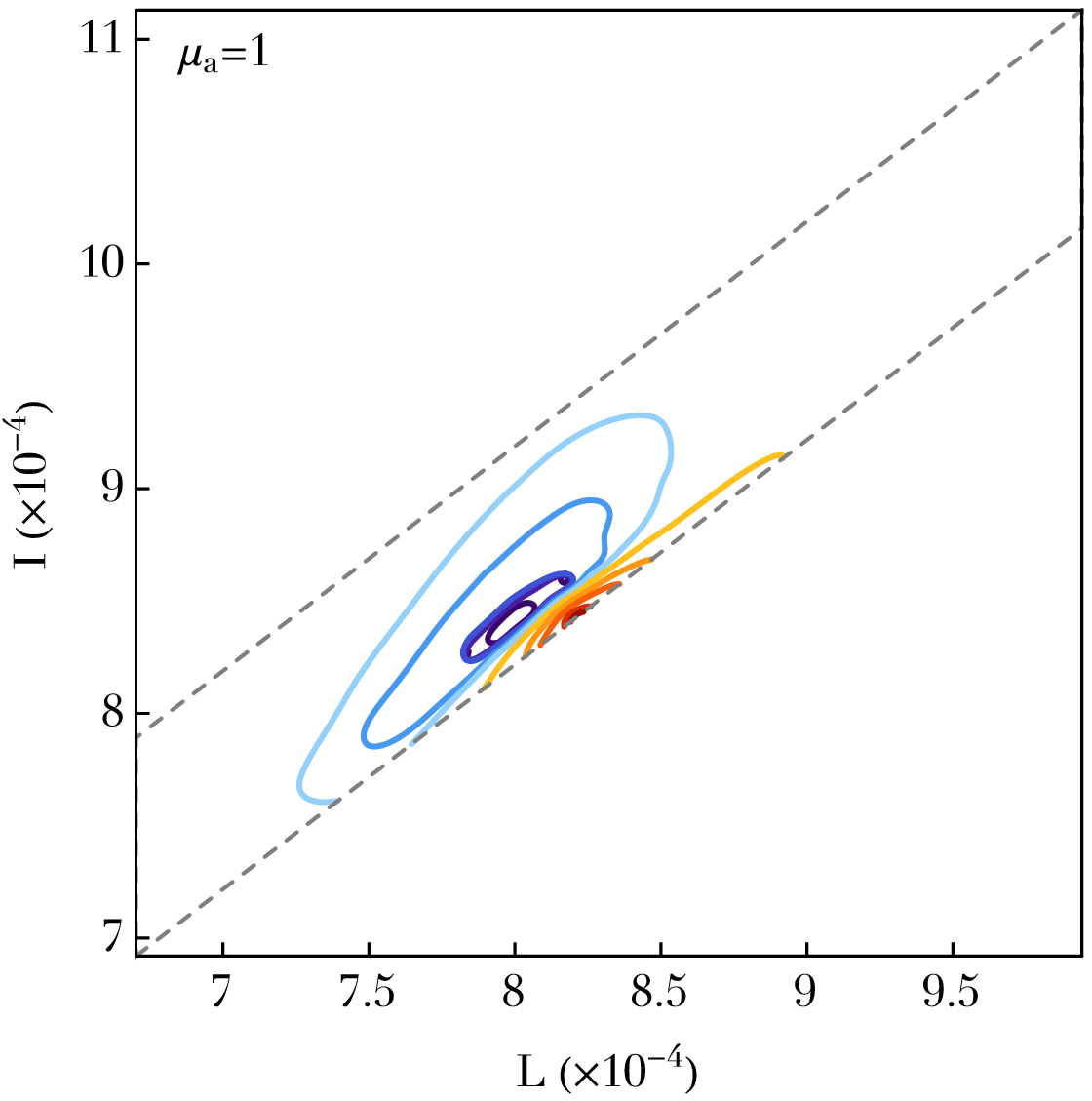,angle=-00,width=0.45\textwidth}
\epsfig{file=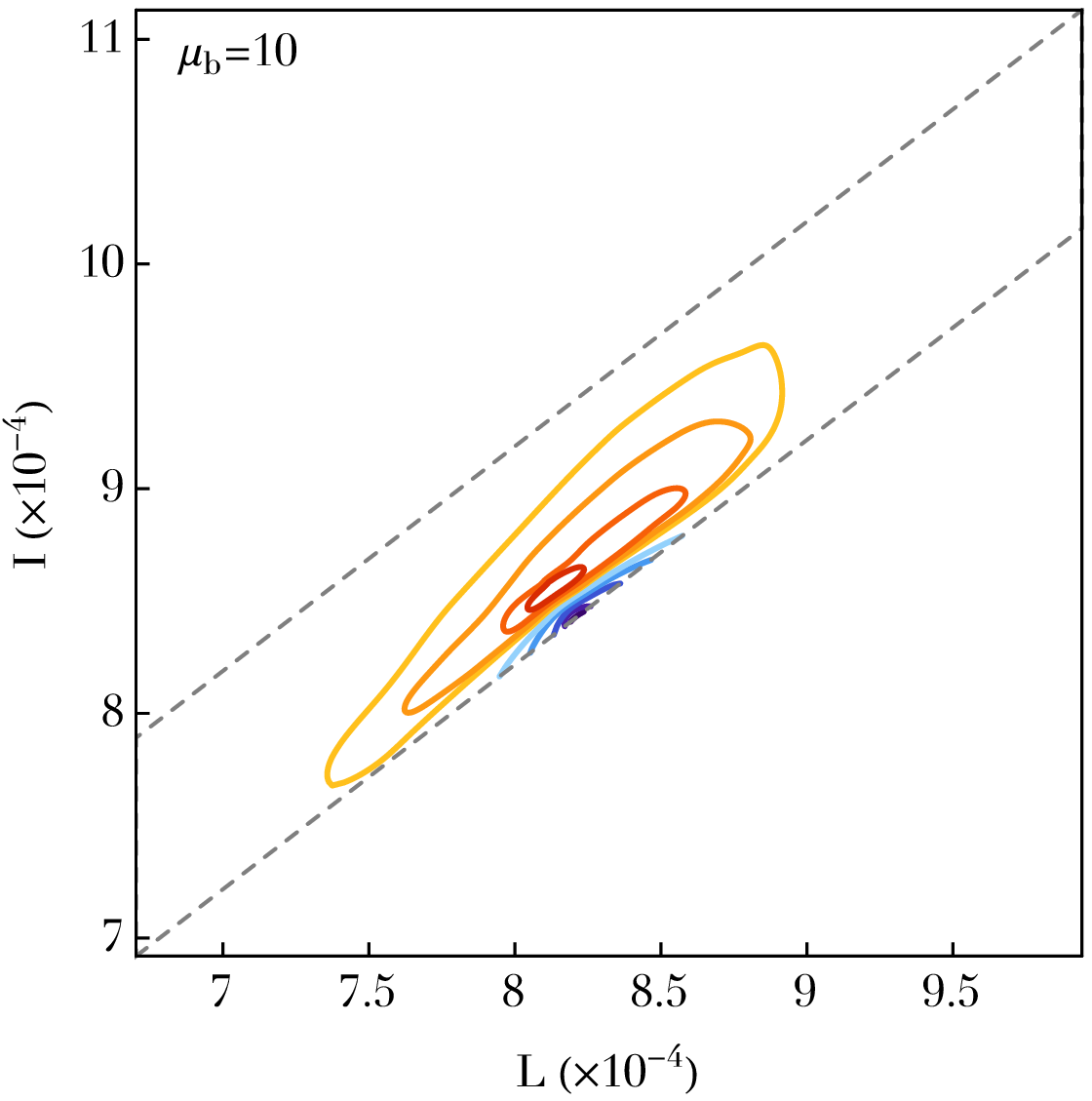,angle=-00,width=0.45\textwidth}
\caption{\small{Illustration of the divergence of the diffusion flux, ${ \text{div} (\bm{\mathcal{F}}_{\rm tot}) }$, predicted by the multi-component Landau equation~\eqref{Landau_Kep_multi} and following the conventions from figure~\ref{figDivFluxSelf}. \textbf{Left panel}. For the population ``$\ra$'' of light particles of individual mass ${ \mu_{\ra} \!=\! 1 }$. The maximum value for the positive blue contours is given by ${ \text{div} (\bm{\mathcal{F}}_{\rm tot}^{\ra})_{\rm max} \!\simeq\! 8 \!\times\! 10^{-13} }$, while the minimum value for the negative red contours reads ${ \text{div} (\bm{\mathcal{F}}_{\rm tot}^{\ra})_{\rm min} \!\simeq\! - 3 \!\times\! 10^{-12} }$. \textbf{Right panel}. For the population ``$\rb$'' of heavy particles of individual mass ${ \mu_{\rb} \!=\! 10 }$. The maximum and minimum values for the contours are given by ${ \text{div} (\bm{\mathcal{F}}_{\rm tot}^{\rb})_{\rm max} \!\simeq\! 3 \!\times\! 10^{-12} }$ and ${ \text{div} (\bm{\mathcal{F}}_{\rm tot}^{\rb})_{\rm min} \!\simeq\! - 10^{-12} }$. Particles are initially distributed according to similar DFs, but undergo a mass segregation on secular timescales. Light particles get larger eccentricity (smaller $L$), while heavy particles circularise (larger $L$).
}}
\label{figDivFluxMulti}
\end{center}
\end{figure*}

\begin{figure*}[!htbp]
\begin{center}
\epsfig{file=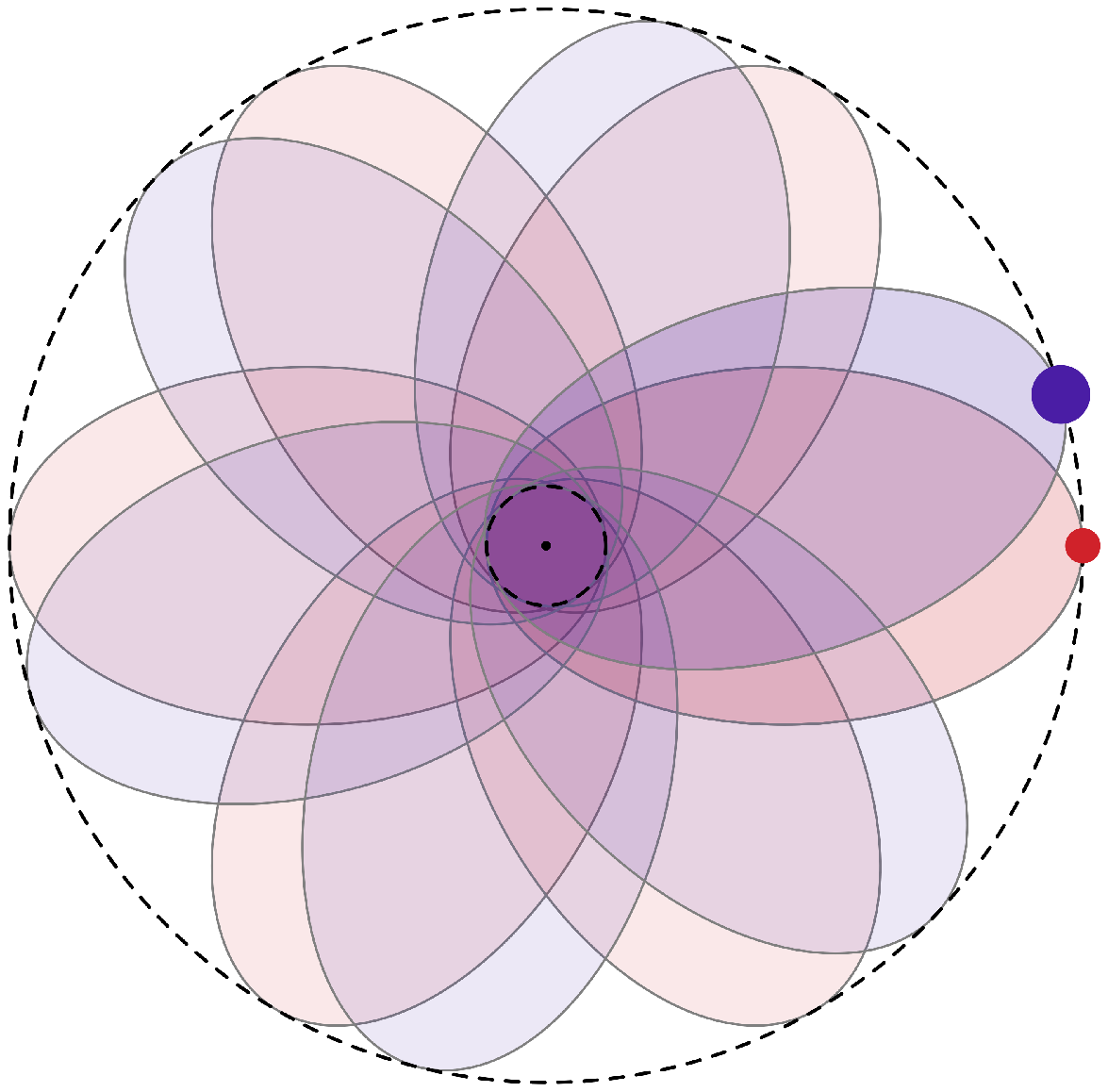,width=0.4\textwidth}
\hspace{0.05\textwidth}
\epsfig{file=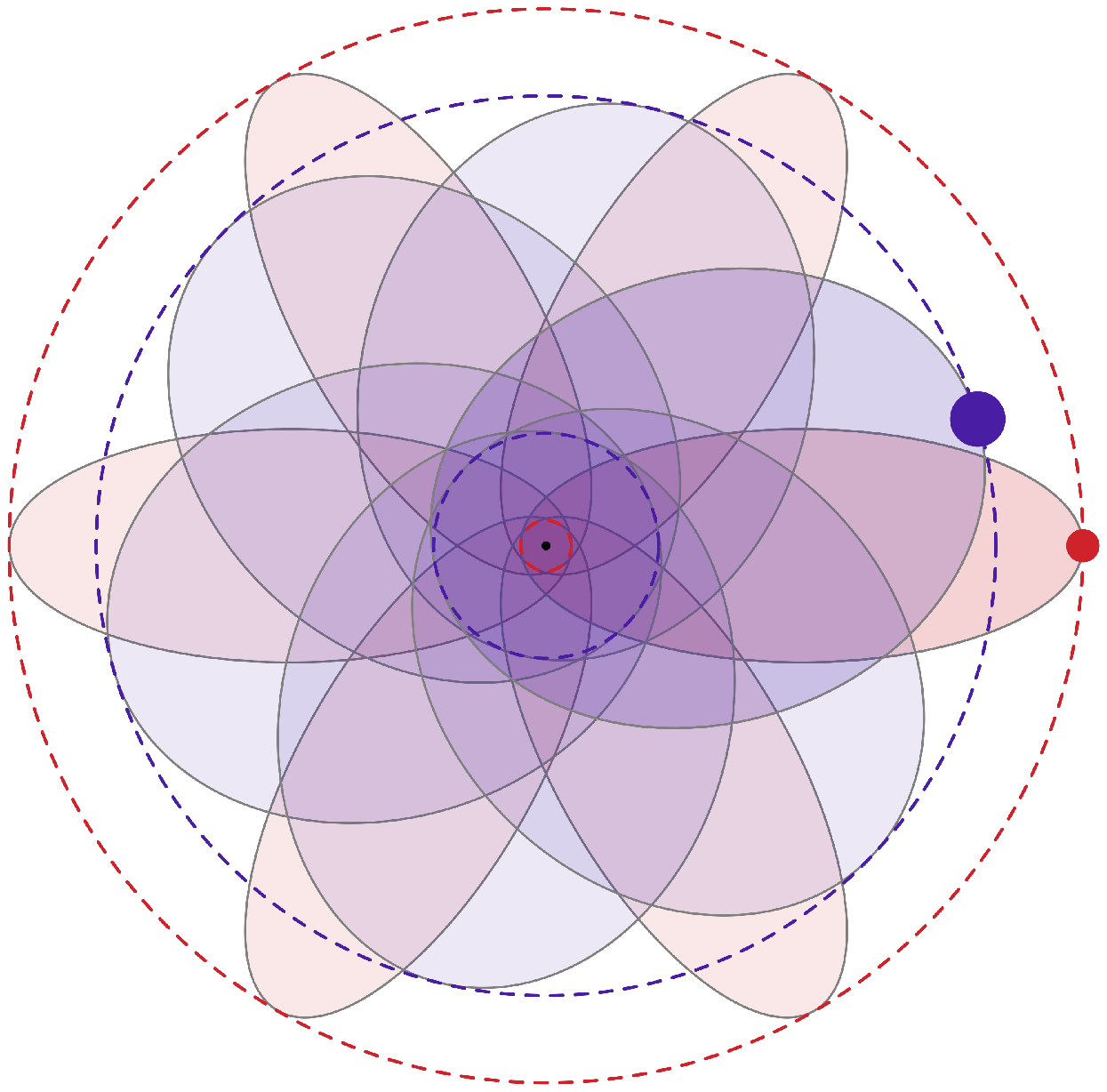,width=0.4\textwidth}
\caption{\small{Illustration in configuration space of the mass segregation of the two different components obtained in figure~\ref{figDivFluxMulti}. Here, the population of red orbits has a lighter individual mass than the blue population. \textbf{Left panel}: Illustration of the initial orbits of the particles, where the blue and red particles have the same semi-major axis and eccentriciy. \textbf{Right panel}: Illustration of the particles' orbits after the resonant mass segregation. During the resonant relaxation, the particles conserve their semi-major axis, but, following figure~\ref{figDivFluxMulti}, the light red particles get larger eccentricities, while the heavy blue particles diffuse towards smaller eccentricities and circularise. Because of this segregation, one can note that red orbits get closer to the central BH, as illustrated by the dashed circles.
}}
\label{figSegregationPhysical}
\end{center}
\end{figure*}

Let us illustrate this multi-component diffusion by considering the exact same disc profile as in section~\ref{sec:discmodel}. However, here it will be assumed that half of the mass of the disc is due to a population of stars whose individual mass is ten times larger than the individual mass considered in the one-component case. Following the units from equation~\eqref{mass_BH_disc}, the two populations ``$\ra$'' and ``$\rb$'' are such that
\begin{equation}
M_{\star}^{\ra} = M_{\star}^{\rb} = \frac{M_{\star}}{2} \;\; ; \;\; \mu_{\ra} = 1 \;\; ; \;\; \mu_{\rb} = 10 \, .
\label{assumption_mass}
\end{equation}
One may then reuse the calculations presented in section~\ref{sec:Landausingle} to compute the divergence of the diffusion flux of each of the two populations ``$\ra$'' and ``$\rb$''. This is illustrated in figure~\ref{figDivFluxMulti}.
In this figure, one can note that the population ``$\ra$'' of light particles tends to diffuse toward larger eccentricities, while the population of ``$\rb$'' of heavy particles diffuses towards smaller eccentricities. This segregation is of particular astrophysical interest in galactic centres in order to investigate how a sub-population of intermediate mass black holes (represented by the heavy particles) may diffuse in these regimes compared to the stellar population. In the present case, the diffusion coefficient from the degenerate Landau equation presented in figure~\ref{figDivFluxMulti} predicts that the heavy population circularise as a result of the self-induced resonant relaxation.
The mass segregation observed in figure~\ref{figDivFluxMulti} has a direct counterpart in configuration space, as illustrated in figure~\ref{figSegregationPhysical}.

In closing, let us briefly recover the mass segregation observed in figure~\ref{figDivFluxMulti} by computing the initial rate of change of the mean angular momentum of each population. Defining
\begin{equation}
\big< L_{\ra} \big> = \!\! \int \!\! \rd \bm{J} \, \Fsa (\bm{J}) \, L \, ,
\label{definition_mean_L}
\end{equation}
and following equation~\eqref{Landau_Kep_disc_multi_tot_AD}, one has
\begin{equation}
\frac{\rd \big< L_{\ra} \big>}{\rd t} = - \eta_{\ra} \!\! \int \!\! \rd \bm{J} \, A^{\rm tot} (\bm{J}) \, \Fsa (\bm{J}) - \!\! \int \!\! \rd \bm{J} \, D^{\rm tot} (\bm{J}) \, \frac{\partial \Fsa}{\partial L} \, .
\label{dL_dt}
\end{equation}
Thanks to equation~\eqref{calculation_Atot_Dtot}, the value of ${ \rd \big< L_{\ra} \big> / \rd t }$ at the initial time is given by
\begin{align}
\frac{\rd \big< L_{\ra} \big>}{\rd t} \bigg|_{0} = & \, - \frac{\mu_{\ra}}{M_{\star}} \frac{M_{\star}}{\mu_{\star}} \frac{M_{\star}^{\ra}}{M_{\star}} \!\! \int \!\! \rd \bm{J} \, A (\bm{J}) \, F_{\star} (\bm{J})  \nonumber
\\
& \, - \frac{M_{\star}^{\ra} \mu_{\ra} \!+\! M_{\star}^{\rb} \mu_{\rb}}{M_{\star} \mu_{\star}} \frac{M_{\star}^{\ra}}{M_{\star}} \!\! \int \!\! \rd \bm{J} \, D (\bm{J}) \, \frac{\partial F_{\star}}{\partial L} \, .
\label{dL_dt_II}
\end{align}
The disc's total angular momentum being conserved~\citep{SridharTouma2016III}, one has
\begin{equation}
\!\! \int \!\! \rd \bm{J} \, A (\bm{J}) \, F_{\star} (\bm{J}) + \!\! \int \!\! \rd \bm{J} \, D (\bm{J}) \, \frac{\partial F}{\partial L} = 0 \, ,
\label{relation_A_D}
\end{equation}
so that equation~\eqref{dL_dt_II} can finally be rewritten as
\begin{equation}
\frac{\rd \big< L_{\ra} \big>}{\rd t} \bigg|_{0} = \frac{M_{\star}^{\ra} M_{\star}^{\rb}}{M_{\star} M_{\star} \mu_{\star}} \, ( \mu_{\ra} \!-\! \mu_{\rb} ) \, \!\! \int \!\! \rd \bm{J} \, D (\bm{J}) \, \frac{\partial F_{\star}}{\partial L} \, .
\label{dL_dt_final}
\end{equation}
Following figure~\ref{figDFstar}, let us assume that ${ \partial F_{\star} / \partial L \!>\! 0 }$ (which is true in most of action space). The diffusion coefficient ${ D (\bm{J}) }$ being always positive, one has
\begin{equation}
\!\! \int \!\! \rd \bm{J} \, D (\bm{J}) \, \frac{\partial F_{\star}}{\partial L} > 0 \, .
\label{sign_D_integral}
\end{equation}
As a consequence, for ${ \mu_{\ra} \!<\! \mu_{\rb} }$, one has ${ \rd \big< L_{\ra} \big> / \rd t |_{0} \!<\! 0 }$ and ${ \rd \big< L_{\rb} \big> / \rd t |_{0} \!>\! 0 }$. Equation~\eqref{dL_dt_final} therefore predicts that as a result of resonant relaxation, the light particles will see a decrease in their mean angular momentum (i.e. an increase in eccentricity), while the heavy particles will see an increase in their mean angular momentum (i.e. a decrease in eccentricity). This corresponds to the segregation observed in figure~\ref{figDivFluxMulti}. Let us finally emphasise that except for specific cases (e.g., here ${ \partial F_{\star} / \partial L \!>\! 0 \,,\, \forall L }$), it remains difficult to predict a priori the direction of mass segregation for other arbitrary initial conditions, as the calculation of the Landau diffusion fluxes from equation~\eqref{Landau_Kep_disc_multi_AD} is very intricate.

\section{Reaching the Schwarzschild barrier}
\label{sec:LangevinBarrier}

The previous section investigated the self-induced diffusion of the disc's DF as a whole. The long-term self-consistent diffusion of this DF is then described by the degenerate Landau equation~\eqref{Landau_Kep_disc}, which is quadratic in the system's DF. Instead of describing the evolution of the disc's DF as a whole, it is of interest to follow the stochastic evolution of arbitrary individual stellar wires, perturbed by the ${1/N}$ noise due to the disc. This would allow us for instance to investigate the impact of the stellar disc on the evolution of stars or intermediate mass black holes in the vicinity of the SMBH. Such stochastic dynamics are captured by a Langevin equation, as described below. In this context the quasi-Keplerian disc will be treated as a bath, so that its mean DF, $\Fs$, does not evolve on the relevant timescale.

\subsection{The stochastic Langevin equation}
\label{sec:Langevinsimple}

 Let us consider a given test star, and represent its statistics by the probability distribution function (PDF), $P$. This PDF describes the stochastic dynamical evolution of individual test wires driven by the ${1/N}$ noise of the disc (the ``bath''). $P$ obeys a Fokker-Planck equation~\citep[and references therein]{Heyvaerts2017} reading
\begin{equation}
\frac{\partial P}{\partial \tau} = \frac{\partial }{\partial L} \bigg[ A (\bm{J}) \, P (\bm{J}) + D (\bm{J}) \, \frac{\partial P}{\partial L} \bigg] \, ,
\label{FP_disc}
\end{equation}
where the drift and diffusion coefficients, ${ A (\bm{J}) }$ and ${ D (\bm{J}) }$, are induced by the disc, and were introduced in equation~\eqref{Landau_Kep_disc_AD}.\footnote{Equation~\eqref{FP_disc} can also straightforwardly be rewritten under the traditional form~\citep{BinneyTremaine2008}
\begin{equation}
\frac{\partial P}{\partial \tau} = \frac{\partial }{\partial L} \bigg[ \!-\! D^{(1)} (\bm{J}) \, P (\bm{J}) + \frac{\partial }{\partial L} \bigg[ D^{(2)} (\bm{J}) \, P (\bm{J}) \bigg] \bigg] \, , \nonumber
\end{equation}
where the first- and second-order diffusion coefficients are given by ${ D^{(1)} \!=\! - A \!+\! \partial D / \partial L }$ and ${ D^{(2)} \!=\! D }$. Here, $D^{(1)}$ captures the true friction force, while ${ - A }$ captures the friction force by polarisation~\citep{Chavanis2012,Heyvaerts2017}.} 
In practice, this equation is obtained by replacing $F_\star$ by $P$ in the flux of equation~\eqref{Landau_Kep_disc_AD}.
The corresponding Langevin equation describes the stochastic dynamics of an individual test wire of action ${ \bm{J}_{\rt} \!=\! (L_{\rt} , I_{\rt}) }$~\citep{Risken1996}. It reads
\begin{equation}
\frac{\rd L_{\rt}}{\rd \tau} = h (\bm{J}_{\rt}) + g (\bm{J}_{\rt}) \, \Gamma (\tau) \;\; ; \;\; \frac{\rd I_{\rt}}{\rd t} = 0 \, .
\label{Langevin_disc}
\end{equation}
Equation~\eqref{Langevin_disc} introduces the ${ 1D }$ Langevin coefficients ${ h (\bm{J}_{\rt}) }$ and ${ g (\bm{J}_{\rt}) }$ defined as
\begin{equation}
h = - A + \frac{\partial D}{\partial L} - \sqrt{D} \frac{\partial \sqrt{D}}{\partial L} = - A + \frac{1}{2} \frac{\partial D}{\partial L} \;\;\; ; \;\;\; g = \sqrt{D} \, .
\label{definition_h_g_disc}
\end{equation}
Finally, equation~\eqref{Langevin_disc} also introduces a Gaussian white noise ${ \Gamma (\tau) }$, whose statistics obeys
\begin{equation}
\big< \Gamma (\tau) \big> = 0 \;\; ; \;\; \big< \Gamma (\tau) \, \Gamma (\tau') \big> = 2 \delta_{\rD} (\tau \!-\! \tau') \, .
\label{noise_Gamma_disc}
\end{equation}
As expected, the stochastic evolution equations~\eqref{Langevin_disc} allows only for diffusion in the ${L_{\rt}-}$direction, while the fast action $L_{\rt}$ of the wire remains conserved during the resonant relaxation.
This stochastic rewriting of the dynamics of a test wire directly connects to the Monte-Carlo approaches considered in~\cite{Madigan2011} and the ${\eta-}$formalism presented in~\cite{BarOrAlexander2014,BarOrAlexander2016}.
The equilibrium solutions of the Fokker-Planck equation~\eqref{FP_disc} are straightforwardly given by
\begin{equation}
P_{\rm eq} (L , I ) = C (I) \, \exp \big[\! - V_{\rm eq} (L , I) \big] \, ,
\label{equilibrium_FP}
\end{equation}
where ${ C (I) }$ is an arbitrary function, and where the potential ${ V_{\rm eq} (L , I) }$ is imposed by the bath and is given by the primitive
\begin{equation}
V_{\rm eq} (L , I) = \!\! \int \!\! \rd L \, \frac{A (I , L)}{D (I , L)} \, .
\label{def_pot_equilibrium}
\end{equation}
If one considers a test particle evolving in Boltzmann bath as given by equation~\eqref{equilibrium_Boltzmann}, the Fokker-Planck equation~\eqref{FP_disc} takes the simpler form
\begin{equation}
\frac{\partial P}{\partial \tau} = \frac{\partial }{\partial L} \bigg[ D (\bm{J}) \, \bigg\{ \frac{\partial P}{\partial L} + (\beta \Omega^{\rs} (\bm{J}) \!-\! \gamma) \, P (\bm{J}) \bigg\} \bigg] \, ,
\label{FP_Boltzamnn_FP}
\end{equation}
thanks to the Einstein relation ${ A (\bm{J}) \!=\! (\beta \Omega^{\rs} (\bm{J}) \!-\! \gamma) \, D (\bm{J}) }$ satisfied by the drift and diffusion coefficients. In the high temperature limit, ${ \beta \!\to\! 0 }$, the Einstein relation becomes ${ A (\bm{J}) \!=\! - \gamma D (\bm{J}) }$. Finally, for an isotropic bath, ${ F_{\star} \!=\! F_{\star} (I) }$, following equation~\eqref{definition_A_D_generic}, the drift coefficients exactly vanish, i.e. ${ A (\bm{J}) \!=\! 0 }$. The associated Fokker-Planck equilibrium states from equation~\eqref{equilibrium_FP} are then also isotropic and read ${ P_{\rm eq} (L , I) \!=\! C (I) }$.

\subsection{Diffusion of an eccentric particle}
\label{sec:LangevinMono}

In the context of the so-called last parsec problem, 
relying on the stochastic Langevin equations~\eqref{Langevin_disc}, let us investigate how a given test wire diffuses in the vicinity of the central BH under the effect of the noise due to the discrete quasi-Keplerian disc. This section will show in particular how the diffusion of this test particle strongly quenches as it reaches large eccentricity, a phenomenon called the ``Schwarzschild barrier'', first observed in~\cite{Merritt2011} in the context of ${3D}$ quasi-Keplerian systems.

Following figure~\ref{figDiscLangevin}, let us therefore consider a test particle of individual mass ${ \mu_{\rt} \!=\! \mu_{\star} }$ and of fast action ${ I_{\rt} (a_{\rt}) \!=\! I_{\rt} (10^{2.5}) }$, where the fast action $I_{\rt}$ and the associated semi-major axis $a_{\rt}$ are directly related by equation~\eqref{definitions_a_e_eta}.
\begin{figure}[!htbp]
\begin{center}
\epsfig{file=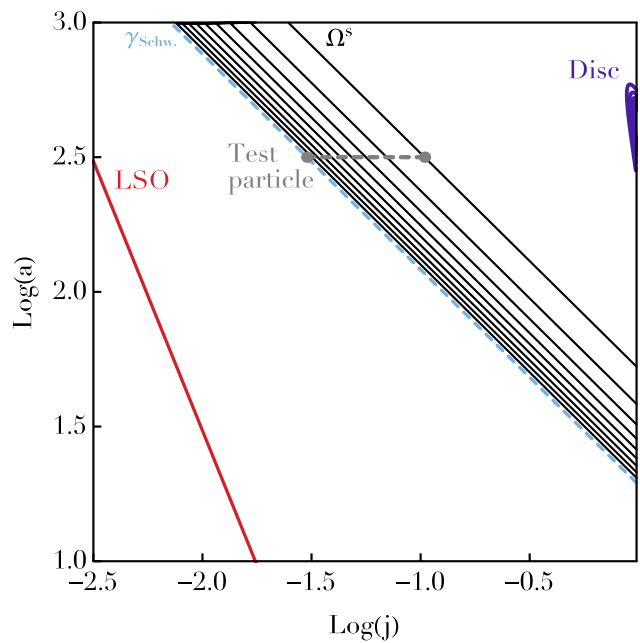,angle=-00,width=0.45\textwidth}
\caption{\small{Illustration of the diffusion of an individual test wire in the ${ (j,a) \!=\! (L/I , I^{2}/(G M_{\bullet}))-}$space. Because of the adiabatic conservation of the fast action $I$, particles diffuse on horizontal lines. The red line corresponds to the last stable orbit, ${ a_{\rm LSO} (j) \!=\! R_{\rm g} (4/j)^{2} }$, with ${ R_{\rm g} \!=\! G M_{\bullet} / c^{2} }$~\citep{BarOrAlexander2016}. The contours of the disc's DF, $\Fs$, introduced in equation~\eqref{DF_disc} are represented by the blue lines. The background lines correspond to some of the level lines of the precession frequency ${ \Omega^{\rs} \!=\! \Omega^{\rs}_{\rm self} \!+\! \Omega^{\rs}_{\rm rel} }$, which are dominated by the relativistic precession $\Omega^{\rs}_{\rm rel}$ for such eccentric orbits. These contours are computed for a retrograde test star, and are therefore associated with negative precession frequencies. They are spaced linearly between the maximum and the minimum precession frequency in the region of the disc, which are dominated by the self-consistent precession $\Omega^{\rs}_{\rm self}$. The dashed gray line corresponds to the segment along which the drift and diffusion coefficients for the test particle are computed in figure~\ref{figDriftDiffLangevin}. (Recall that the test star is assumed to be retrograde, i.e. ${ L_{\rt} \!<\! 0 }$, but for clarity, it is represented on the same diagram.) The cyan line illustrates the location of the Schwarzschild barrier, $\gamma_{\rm Schw.}$, for retrograde test stars defined in equation~\eqref{def_barrier}. Retrograde test particles to the left of this barrier will precess too fast to resonate with this disc, see figures~\ref{figDriftDiffLangevin} and~\ref{fighgLangevin}. Such particles do not undergo any resonant relaxation, and can only diffuse as a result of additional diffusion mechanisms, such as ${2-}$body non-resonant relaxation.
}}
\label{figDiscLangevin}
\end{center}
\end{figure}
Any wire in the system undergoes two simultaneous precessions, given by ${ \Omega^{\rs} \!=\! \Omega^{\rs}_{\rm self} \!+\! \Omega^{\rs}_{\rm rel} }$. As emphasised in~\cite{KocsisTremaine2011}, one can note that the self-consistent mass precession frequencies $\Omega^{\rs}_{\rm self}$ induced by the disc are retrograde precessions (i.e. ${ \Omega^{\rs}_{\rm self} \!<\! 0 }$ for ${ L \!>\! 0 }$), while the relativistic precession frequencies $\Omega^{\rs}_{\rm rel}$ are prograde precessions (i.e. ${ \Omega^{\rs}_{\rm rel} \!>\! 0 }$ for ${ L \!>\! 0 }$). Because the mass precession dominates the precessions in the vicinity of the disc, a wire located within the disc region will undergo a retrograde precession, while a wire located close to the central BH will mainly precess as a result of the relativistic precessions and therefore will undergo a prograde precession. Note that the resonance condition present in the Landau equation~\eqref{Landau_Kep_disc} is sign-dependent, i.e. requests to exactly match the precession of the resonating particles so that ${ \Omega^{\rs} (\bm{J}_{1}) \!=\! \Omega^{\rs} (\bm{J}_{2}) }$. As a consequence, for a test wire located close to the central BH to be able to resonate with a disc composed only of prograde orbits (i.e. ${ L \!>\! 0 }$), this test wire has to be retrograde (i.e. ${ L_{\rt} \!<\! 0 }$), as we will now assume. Should the test wire in the central wire be also prograde, no efficient resonant couplings would be permitted by the Landau equation~\eqref{Landau_Kep_disc} and the associated diffusion would tend to 0. Let us note that this requirement on the central test wire direction of rotation arises from the additional constraints associated with the disc's geometry. For a $3D$ spherical quasi-Keplerian systems, the Landau equation~\eqref{Landau_Kep_disc} would allow for additional resonances. This will be the subject of a future work.

\begin{figure*}[!htbp]
\begin{center}
\epsfig{file=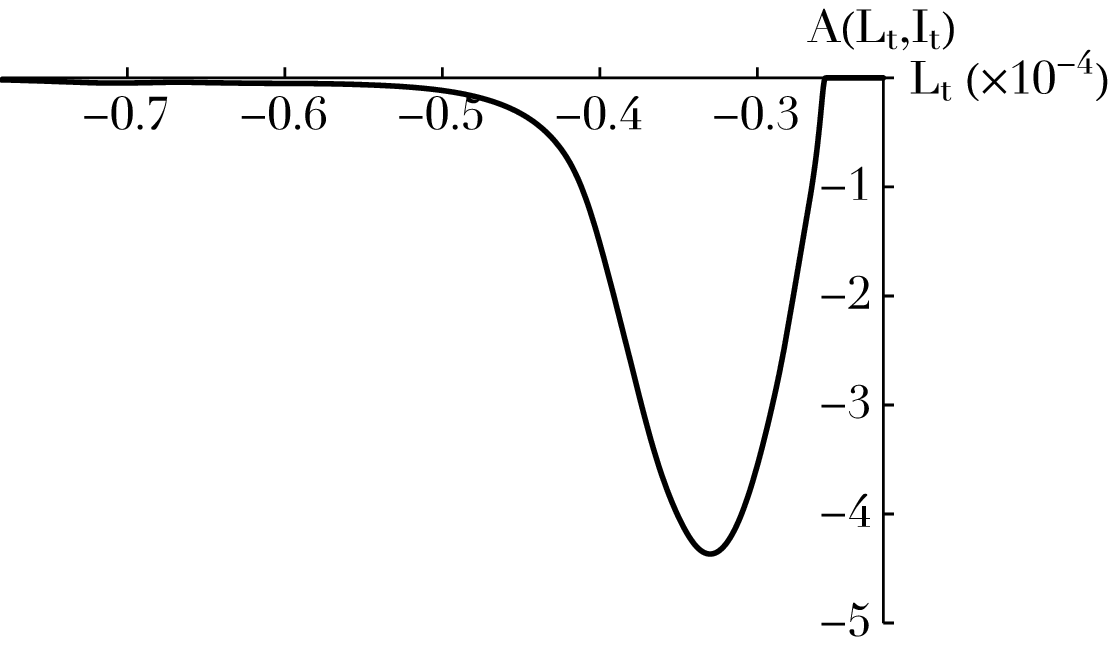,angle=-00,width=0.45\textwidth}
\epsfig{file=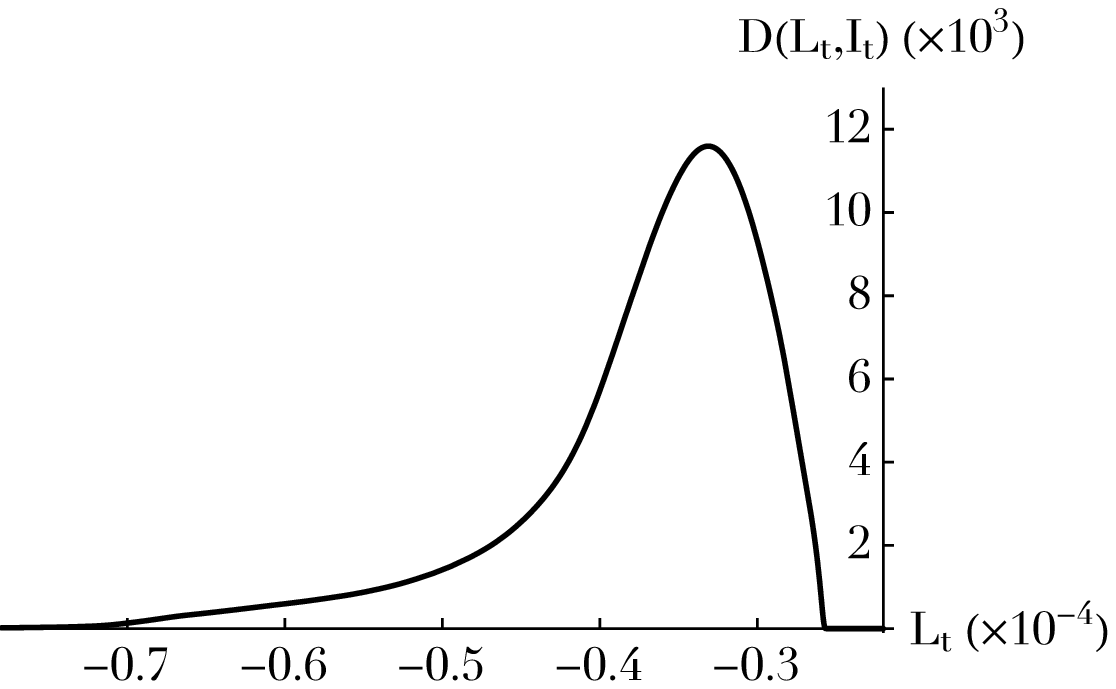,angle=-00,width=0.45\textwidth}
\caption{\small{Illustration of the drift and diffusion coefficients for a retrograde test orbit diffusing in the inner region of the system along the gray dashed line, ${ I_{\rt} \!=\! \text{cst.} }$, identified in figure~\ref{figDiscLangevin}. \textbf{Left panel}. Illustration of the drift coefficient ${ L_{\rt} \!\mapsto\! A (L_{\rt} , I_{\rt}) }$, as introduced in equation~\eqref{FP_disc}. \textbf{Right panel}. Illustration of the diffusion coefficient ${ L_{\rt} \!\mapsto\! D (L_{\rt} , I_{\rt}) }$, as introduced in equation~\eqref{FP_disc}. As the test particle gets closer to the centre of the system, the drift and diffusion coefficients tend to $0$: this is the Schwarzschild barrier, which prevents individual stars to diffuse closer to the central BH, as a sole result of resonant relaxation. The quenching of the resonant diffusion is very abrupt in razor-thin discs, as a result of the limitation to ${1\!:\!1}$ resonance in the degenerate Landau equation~\eqref{Landau_Kep_disc}. This is specific to the razor-thin geometry.
}}
\label{figDriftDiffLangevin}
\end{center}
\end{figure*}

As shown in figure~\ref{figDiscLangevin}, one may then study the stochastic diffusion of such a retrograde test wire along the gray segment where it may resonate with the outer quasi-Keplerian disc. This is illustrated in figure~\ref{figDriftDiffLangevin} where the drift and diffusion coefficients associated with the diffusion of this test wire are 
computed following equation~\eqref{FP_disc}. Recall that because the test star is assumed to be retrograde, one has ${ L_{\rt} \!<\! 0 }$.
In figure~\ref{figDriftDiffLangevin}, one can note that for ${ | L_{\rt} | \!\lesssim\! 2.7 \!\times\! 10^{3} }$, the drift and diffusion coefficients tend to 0. This is the Schwarzschild barrier. Particles of high eccentricity, i.e. particles which get close to the central BH undergo a large relativistic precession. For eccentricities large enough, this relativistic precession gets so large that it prevents any coupling between the test wire and wires within the disc. The resonant relaxation stops. For a razor-thin disc, the quenching is very abrupt as for low enough ${ | L_{\rt} | }$, the drift and diffusion coefficients tend to 0. This is a direct consequence of the Landau equation~\eqref{Landau_Kep_disc}, which for razor-thin discs, only allows for ${1\!:\!1}$ resonances. For ${3D}$ systems, the geometric constraint on the allowed resonances weakens. Higher-order resonances, while associated with weaker coupling factors, are allowed by the kinetic equation, so that the quenching of the resonant relaxation in the vicinity of the Schwarzschild barrier is expected to be less abrupt compared to what has been obtained in figure~\ref{figDriftDiffLangevin}. In practice, this suppression of the diffusion is tempered by simple two-body relaxation, not accounted for in the present orbit-averaged diffusion. This provides an additional mechanism allowing stars to diffuse closer to the BH, once resonant relaxation becomes inefficient. As demonstrated in~\cite{BarOrAlexander2016}, the effects of resonant relaxation are limited to regions well away of the loss cone, so that the dynamics of stars' accretion is only moderately affected by the presence of resonances.

\begin{figure*}[!htbp]
\begin{center}
\epsfig{file=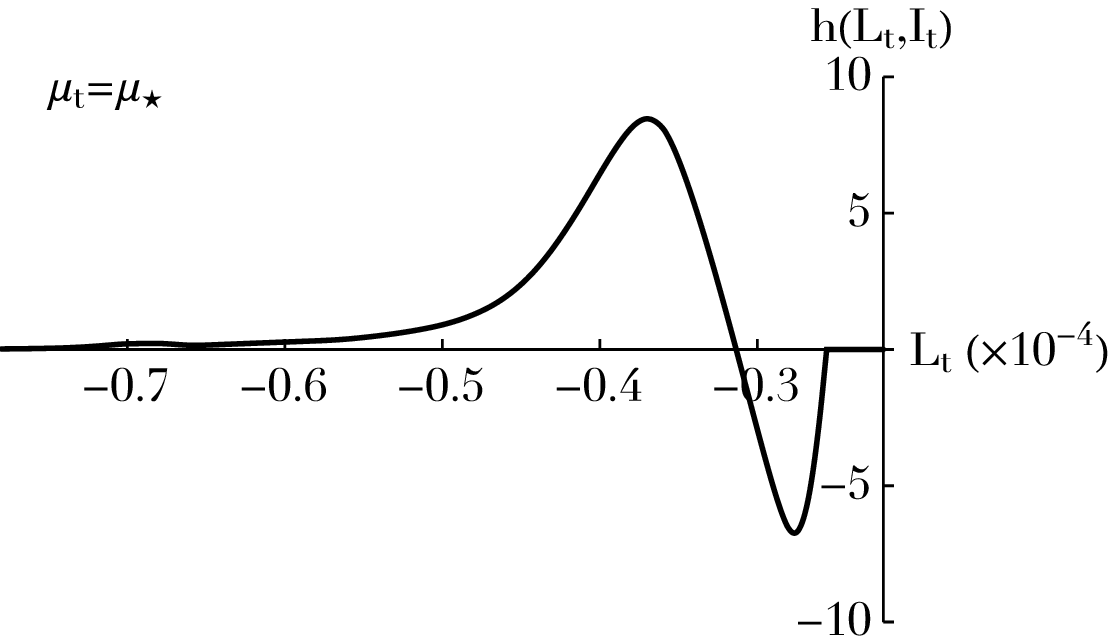,angle=-00,width=0.45\textwidth}
\epsfig{file=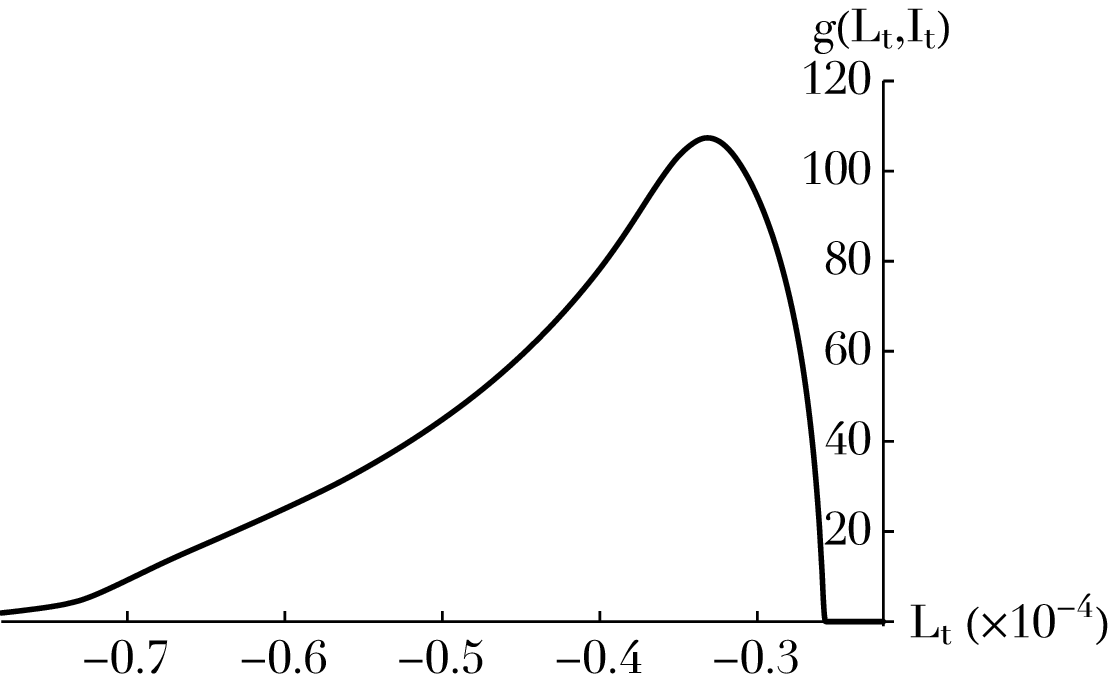,angle=-00,width=0.45\textwidth}
\caption{\small{Illustration of the stochastic Langevin coefficients associated with the drift and diffusion coefficients shown in figure~\ref{figDriftDiffLangevin}. \textbf{Left panel}: Illustration of the Langevin drift coefficient ${ L_{\rt} \!\mapsto\! h (L_{\rt} , I_{\rt}) }$. Following equation~\eqref{Langevin_disc}, this coefficient gives the mean direction of diffusion for a given location in action space. \textbf{Right panel}: Illustration of the Langevin diffusion coefficient ${ L_{\rt} \!\mapsto\! g (L_{\rt} , I_{\rt}) }$. This coefficient describes the jitter of particles around the mean flow given by $h$. In particular, it allows particles to stochastically penetrate the barrier.
}}
\label{fighgLangevin}
\end{center}
\end{figure*}

Following the computation of the drift and diffusion coefficients in figure~\ref{figDriftDiffLangevin}, one may then rely on equation~\eqref{definition_h_g_disc} to estimate the Langevin coefficients, $h$ and $g$, characterising the stochastic diffusion of the test wire. These coefficients are illustrated in figure~\ref{fighgLangevin}.
As already noted in figure~\ref{figDriftDiffLangevin}, the Langevin coefficients tend to 0 for ${ |L_{\rt}| \!\lesssim\! 2.7 \!\times\! 10^{3} }$, which corresponds to the Schwarzschild barrier. In the Langevin equation~\eqref{definition_h_g_disc}, the coefficient $h$ corresponds to the drift coefficient and describes the mean deterministic motion followed by the test particle. Here, it is negative right before the barrier, so that retrograde test particles in the vicinity of the barrier diffuse in average towards larger ${ |L_{\rt} |}$, i.e. towards smaller eccentricities. In equation~\eqref{definition_h_g_disc}, the coefficient $g$ is associated with the stochastic diffusion of the test particle. It describes the jitter of the test particle around the mean flow due to $h$. On the long-term, it can allow particles to stochastically penetrate the diffusion barrier. Finally, while the drift coefficient ${ - A (\bm{J}_{\rt}) }$ is always positive in figure~\ref{figDriftDiffLangevin}, the contributions from the diffusion coefficient in equation~\eqref{definition_h_g_disc} lead to a Langevin drift coefficient $h$ taking both positive and negative values in figure~\ref{fighgLangevin}.

Figures~\ref{figDriftDiffLangevin} and~\ref{fighgLangevin} recover the diffusion barrier for a retrograde test wire of fast action $I_{\rt}$. The location of this quenching of the resonant diffusion can be interpreted as given by the value of the slow action $L_{\rm Schw.}$, such that
\begin{equation}
\Omega^{\rs} (L_{\rm Schw.} , I_{\rt}) \simeq \Omega^{\rm max}_{\rm disc} \, ,
\label{pos_barrier}
\end{equation}
where $\Omega^{\rm max}_{\rm disc}$ is the typical maximum precession frequency in the disc region, i.e. the maximum value of ${ \Omega^{\rm s} }$ in figure~\ref{figContoursOmega}. For a retrograde test wire such that ${ | L_{\rt} | \!\lesssim\! L_{\rm Schw.} }$, its relativistic Schwarschild precession makes it precess too fast to allow for a resonant coupling with the disc and the diffusion quenches. Following the criteria from equation~\eqref{pos_barrier}, the location of the barrier for retrograde test stars is then given in action space by the curve $\gamma_{\rm Schw.}$, such that
\begin{equation}
\gamma_{\rm Schw.} = \bigg\{ (L_{\rt} , I_{\rt}) \; \big| \; \Omega^{\rs } (L_{\rt} , I_{\rt}) = \Omega^{\rm max}_{\rm disc} \bigg\} \, .
\label{def_barrier}
\end{equation}
The location of this barrier is illustrated in figure~\ref{figDiscLangevin}, where it is given by the left-most level contours of $\Omega^{\rm s}$. Retrograde test particles below this barrier are precessing too fast to resonate anymore with the disc. Different retrograde test particles having different fast actions $I_{\rt}$ will therefore see their stochastic diffusion quench for different values of their slow action $L_{\rt}$.

\begin{figure}[!htbp]
\begin{center}
\epsfig{file=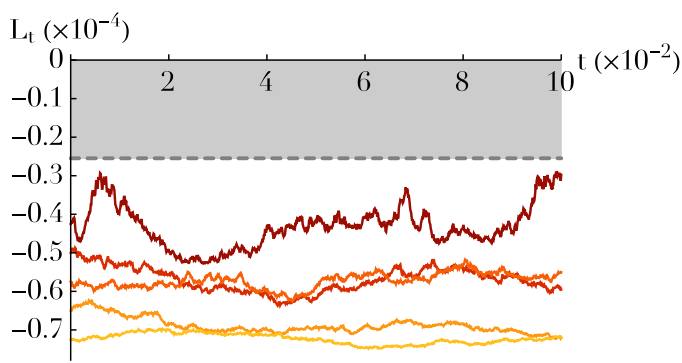,angle=-00,width=0.48\textwidth}
\caption{\small{Illustration of the stochastic motion, ${ t \!\mapsto\! L_{\rt} (t) }$, of a retrograde test star of mass ${ \mu_{\rt} \!=\! \mu_{\star} }$ for different initial conditions. The trajectory of the star is described by the Langevin equation~\eqref{Langevin_disc}, with the Langevin coefficients $h$ and $g$ obtained in figure~\ref{fighgLangevin}. Because these coefficients tend to $0$ for low enough angular momentum $ ( |L_{\rt}| \!\lesssim\! 2.7 \!\times\! 10^{3}) $, retrograde test stars cannot diffuse closer to the BH. This quenching of the resonant diffusion in the inner regions of the system is associated with the Schwarzshild barrier and is illustrated with the gray region.
}}
\label{figStochasticTrajectories}
\end{center}
\end{figure}

Having computed the Langevin coefficients $h$ and $g$ in figure~\ref{fighgLangevin}, it is then straightforward to integrate the Langevin equation~\eqref{Langevin_disc} forward in time. Such realisations are illustrated in figure~\ref{figStochasticTrajectories},
which shows again that particles cannot diffuse below the Schwarzschild barrier. These evolution equations share some similarities with the equations of motions of individual stars. However, the significant gain of this framework is that it directly describes the stochastic motion of Keplerian wires, so that the Keplerian motion of stars along their quasi-Keplerian ellipses does not have to be resolved anymore. This allows for much larger timesteps in equation~\eqref{Langevin_disc}, which are orders of magnitude larger than those required to solve the individual trajectories of stars. Relativistic effects and the associated post-Newtonian corrections are also effortlessly accounted for.

\begin{figure}[!htbp]
\begin{center}
\epsfig{file=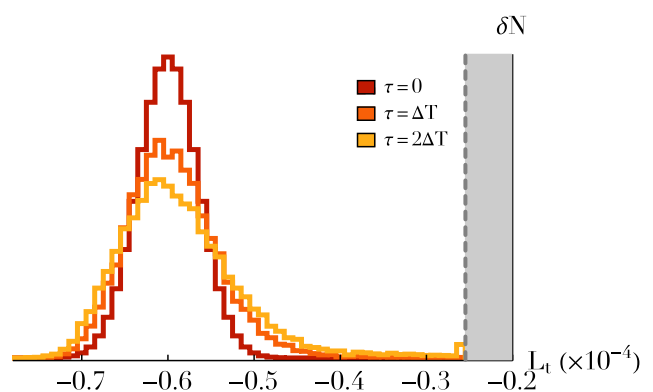,angle=-00,width=0.45\textwidth}
\caption{\small{Illustration of the diffusion of a population of retrograde test wires of individual mass ${ \mu_{\rt} \!=\! \mu_{\star} }$ as a function of time. The evolution of each star is driven by the Langevin equation~\eqref{Langevin_disc}. The initial PDF of the population is represented by the red histogram, while the colored histograms describe the statistics of the population after a time ${ \Delta T \!=\! 200 }$ and ${ 2 \Delta T }$. Solving the dynamics of this population via the Langevin equation~\eqref{Langevin_disc} allows for the integration forward in time of the Fokker-Planck equation~\eqref{FP_disc}, which describes the diffusion of the test particles' PDF as a whole, without resorting to direct ${N-}$body simulations.
}}
\label{figHistogramSelf}
\end{center}
\end{figure}

Not only can one use the Langevin equation~\eqref{Langevin_disc} to describe the evolution of an individual test particle, but also the secular diffusion of a population of wires as a whole. This is illustrated in figure~\ref{figHistogramSelf}, which shows how the long-term diffusion of the PDF of a population of retrograde test particles may also be estimated.
The method followed in figure~\ref{figHistogramSelf} allows indeed for the effective integration forward in time of the Fokker-Planck equation~\eqref{FP_disc}. To do so, one samples the test particle's PDF, $P$, with test particles. The stochastic motion of each test particle is then integrated forward in time via the Langevin equation~\eqref{Langevin_disc} for a time ${ \Delta T }$ that can be much larger than the Keplerian dynamical time of the system. After a time ${ \Delta T }$, the population of test particles is then distributed according to the PDF ${ P (t \!=\! \Delta T) }$, illustrated in figure~\ref{figHistogramSelf}. In this figure, even if the time of integration was short, one can already note that some particles tend to accumulate at the ``Schwarzschild barrier'', where the diffusion quenches.

The sampling method used in figure~\ref{figHistogramSelf} may also be used to integrate forward in time the self-consistent Landau equation~\eqref{Landau_Kep_disc}. To do so, one has to estimate the disc's drift and diffusion coefficients ${ A (\bm{J}) }$ and ${ D (\bm{J}) }$. The disc's initial DF, $F_{\star}$ is then sampled by a finite number of test stars ${ N_{\rm samp.} }$. Assuming temporarily that the drift and diffusion coefficients are frozen, one may then integrate the motion of these ${ N_{\rm samp.} }$ test stars following the Langevin equation~\eqref{Langevin_disc}. This allows for the estimation of ${ P (t \!=\! \Delta T) \!\simeq\! F_{\star} (t \!=\! \Delta T) }$, provided that ${ \Delta T }$ is not too large compared to the timescale of resonant relaxation. Having estimated the disc's new DF at the time ${ \Delta T }$, one may then recompute the new drift and diffusion coefficients of the disc, ${ A ( \bm{J} , \Delta T ) }$ and ${ D (\bm{J} , \Delta T) }$. Sampling once again this new DF with $N_{\rm samp.}$ test stars, one could proceed further: provided that the timestep ${ \Delta T }$ is chosen accordingly, so that the disc's self-consistent drift and diffusion coefficients do not change much on the timescale ${ \Delta T }$, the present step-by-step approach allows therefore for the integration forward in time of the self-consistent Landau equation~\eqref{Landau_Kep_disc}.

\subsection{Resonant dynamical friction on a massive perturber}
\label{sec:Langevinheavy}

The previous section described the stochastic diffusion of an individual test star, whose individual mass is identical to that of the stars forming the discrete quasi-Keplerian disc. Inspired by the multi-component calculations presented in section~\ref{sec:LandauMultidisc}, one could also consider the individual diffusion of a massive perturber whose mass would not be the same as the particles from the discrete bath. Noting the mass of this test perturber as $\mu_{\rt}$ and the individual mass of the particles of the bath as $\mu_{\star}$, the Fokker-Planck equation~\eqref{FP_disc} becomes
\begin{equation}
\frac{\partial P}{\partial \tau} = \frac{\partial }{\partial L} \bigg[ \frac{\mu_{\rt}}{\mu_{\star}} \, A (\bm{J}) \, P (\bm{J}) + D (\bm{J}) \, \frac{\partial P}{\partial L} \bigg] \, ,
\label{FP_disc_heavy}
\end{equation} 
where $P$ is the PDF of the massive perturber. In equation~\eqref{FP_disc_heavy}, the drift and diffusion coefficients, ${ A (\bm{J}) }$ and ${ D (\bm{J}) }$, were already introduced in equation~\eqref{Landau_Kep_disc_AD} and are sourced by the discrete quasi-Keplerian disc. When accounting for a possible different mass for the test particle, the equilibrium solution from equation~\eqref{equilibrium_FP} immediately becomes
\begin{equation}
P_{\rm eq} (L , I) = C (I) \, \exp \big[\! - ( \mu_{\rt} / \mu_{\star} ) V_{\rm eq} (L , I) \big] \, ,
\label{equilibrium_FP_multi}
\end{equation}
where the potential ${ V_{\rm eq} (L , I) }$ was introduced in equation~\eqref{def_pot_equilibrium}.

\begin{figure}[!htbp]
\begin{center}
\epsfig{file=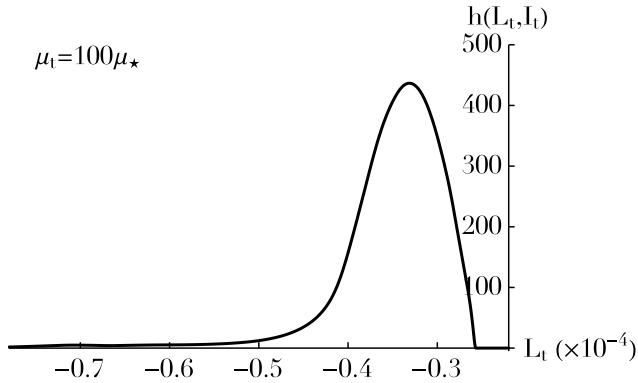,angle=-00,width=0.45\textwidth}
\caption{\small{Illustration of the stochastic Langevin coefficient ${ L_{\rt} \!\mapsto\! h (L_{\rt} , I_{\rt}) }$ associated with the stochastic diffusion of a retrograde massive perturber of mass ${ \mu_{\rt} \!=\! 100 \mu_{\star} }$ along the gray dashed line, ${ I_{\rt} \!=\! \text{cst.} }$, identified in figure~\ref{figDiscLangevin}. The coefficient $g$ associated with the stochastic of this massive perturber is the same as in figure~\ref{fighgLangevin}. Following equation~\eqref{FP_disc_heavy}, one can note that for a massive enough perturber (or for light enough bath particles), one has ${ h (\bm{J}_{\rt}) \!\to\! - (\mu_{\rt} / \mu_{\star}) A (\bm{J}_{\rt}) }$ and ${ g (\bm{J}_{\rt}) \!\to\! 0 }$. This non-vanishing contribution is the friction force by polarisation, which drives dynamical friction.
}}
\label{fighgLangevinMassive}
\end{center}
\end{figure}

Following equation~\eqref{definition_h_g_disc}, one can straightforwardly obtain the Langevin coefficients associated with the Fokker-Planck equation~\eqref{FP_disc_heavy}. They read
\begin{equation}
h = - \frac{\mu_{\rt}}{\mu_{\star}} A + \frac{1}{2} \frac{\partial D}{\partial L} \;\;\; ; \;\;\; g = \sqrt{D} \, .
\label{definition_h_g_disc_heavy}
\end{equation}
In equation~\eqref{definition_h_g_disc_heavy}, one can note that only the Langevin drift coefficient $h$ depends on the mass of the test particle. Figure~\ref{fighgLangevinMassive} illustrates this coefficient for a retrograde massive test particle of mass ${ \mu_{\rt} \!=\! 100 \mu_{\star} }$.
Let us note that the definition from equation~\eqref{definition_A_D_generic} is such that the disc's drift and diffusion coefficients, ${ A (\bm{J}) }$ and ${ D (\bm{J}) }$, satisfy ${ A , D \!\propto\! \mu_{\star} }$. The larger the number of particles in the disc, the slower the diffusion. As a consequence, in the limit of a collisionless bath, i.e. when ${ \mu_{\star} \!\to\! 0 }$, only the drift component remains in equation~\eqref{FP_disc_heavy}. This corresponds to the friction force by polarisation, which does not vanish in the collisionless limit~\citep{Heyvaerts2017}. Following equation~\eqref{definition_h_g_disc_heavy}, one can note that in this collisionless limit only the drift coefficient ${ h (\bm{J}_{\rt}) \!\to\! - (\mu_{\rt} / \mu_{\star}) \, A (\bm{J}_{\rt}) }$ remains in the Langevin equation~\eqref{Langevin_disc}. The evolution of the test particle is fully deterministic and, following equation~\eqref{Langevin_disc}, reads
\begin{equation}
\frac{\rd L_{\rt}}{\rd t} = h (L_{\rt} , I_{\rt}) = - \mu_{\rt} \frac{A}{\mu_{\star}} \, ,
\label{eq_dynamical_friction}
\end{equation}
where, following equation~\eqref{definition_A_D_generic}, ${ A/\mu_{\star} }$ is independent of $\mu_{\star}$. Equation~\eqref{eq_dynamical_friction} is the equation describing dynamical friction. Comparing figures~\ref{fighgLangevin} and~\ref{fighgLangevinMassive}, one can note that for a test particle of individual mass ${ \mu_{\rt} \!=\! 100 \mu_{\star} }$, the Langevin coefficients satisfy ${ g \!\lesssim\! h }$. As a consequence, the evolution of such a heavy particle can be approximated by the deterministic equation~\eqref{eq_dynamical_friction}. Comparing figures~\ref{fighgLangevin} and~\ref{fighgLangevinMassive}, one can also note that for a massive enough retrograde test particle, one has ${ h (L_{\rt}) \!>\! 0 }$ for ${ L_{\rt} \!<\! 0 }$. As a consequence, the dynamical friction undergone by this retrograde massive perturber induces a drift towards smaller ${ | L_{\rt} | }$, i.e. towards higher eccentricities: the orbit of this retrograde massive perturber gets more excentric.

\begin{figure}[!htbp]
\begin{center}
\epsfig{file=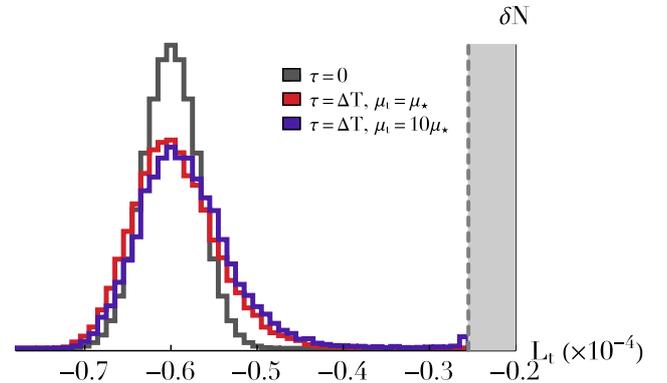,angle=-00,width=0.45\textwidth}
\caption{\small{Illustration of the diffusion of two populations of retrograde test stars of different individual mass. The two populations are initially distributed according to the same PDF, illustrated with the black histogram. The evolution of each test star is described by the Langevin equation associated with the Fokker-Planck equation~\eqref{FP_disc_heavy}. After a time ${ \Delta T \!=\! 200 }$, the PDF of the light popopulation (of individual mass ${ \mu_{\rt} \!=\! \mu_{\star} }$) is given by the red histogram, while the heavy population (of individual mass ${ \mu_{\rt} \!=\! 10 \mu_{\star} }$) follows the PDF given by the blue histogram. Because of the prefactor ${ ( \mu_{\rt} / \mu_{\star} ) }$ present in equation~\eqref{FP_disc_heavy}, populations of different individual mass do not follow the same stochastic motions, and the system undergoes a mass segregation. Light (red) particles tend to become less eccentric and heavy (blue) particles tend to become more eccentric.
}}
\label{figHistogramMulti}
\end{center}
\end{figure}

Expanding on section~\ref{sec:LandauMultidisc}, let us finally investigate the process of mass segregtion using the Langevin formalism. Having already estimated the disc's drift and diffusion coefficients in figure~\ref{figDriftDiffLangevin}, one may now rely on equation~\eqref{definition_h_g_disc_heavy} to compute the Langevin coefficients of populations of retrograde test stars of different individual mass. Figure~\ref{figHistogramMulti} presents the respective diffusion of two populations of retrograde test stars of individual mass ${ \mu_{\rt} \!=\! \mu_{\star} }$ and ${ \mu_{\rt} \!=\! 10 \mu_{\star} }$, distributed initially according to the same PDF.
Figure~\ref{figHistogramMulti} predicts that populations of retrograde test particles of different mass segregate in the vicinity of the disc. The heavier particles will tend towards orbits of smaller angular momentum, i.e. towards more eccentric orbits. One can also note that some heavy particles already tend to accumulate at the ``Schwarzschild barrier'', where resonant diffusion stops.

Figure~\ref{figDivFluxMulti} emphasises that heavy prograde stars in the disc would tend to diffuse towards smaller eccentricities, while figure~\ref{figHistogramMulti} shows that heavy retrograde test stars would preferentially segregate towards higher eccentricities. Let us clarify the origin of this dichotomy. Such differences in the behaviours of prograde and retrograde massive test stars originates from the fact that the razor-thin degenerate Landau equation~\eqref{Landau_Kep_disc} only allows for ${1 \!:\! 1}$ resonances, and that the quasi-Keplerian razor-thin disc considered in equation~\eqref{DF_disc} is only composed of prograde stars. Let us illustrate this property by computing the sign of the friction force by polarisation undergone by a massive perturber, as given by equation~\eqref{eq_dynamical_friction}. Following figure~\ref{figDFstar}, let us assume that ${ \partial F_{\star} / \partial L \!>\! 0 }$ (this is true in most of action space). The expression of the drift coefficients from equation~\eqref{definition_A_D_generic} gives us then that ${ A (\bm{J}) \!\leq\! 0 }$ for all $\bm{J}$, i.e. whatever the sign of $L$, i.e. for both prograde and retrograde stars. As a consequence, the direction of the associated dynamical friction given by equation~\eqref{eq_dynamical_friction} reads
\begin{equation}
\forall \bm{J} \, , \frac{\partial F_{\star} (\bm{J})}{\partial L} > 0 \;\; \Longrightarrow \;\; \frac{\rd L_{\rt}}{\rd t} \bigg|_{\text{Fric.}} \!\!\! > 0 \, .
\label{sign_friction}
\end{equation}
Note that the result from equation~\eqref{sign_friction} is independent of the sign of the angular momentum $L_{\rt}$ of the considered test star. As a consequence, for a prograde test star (i.e. ${ L_{\rt} \!>\! 0 }$), the friction force leads to a diffusion towards larger ${ |L_{\rt}| }$, i.e. towards smaller eccentricities, while for a retrograde test star (i.e. ${ L_{\rt} \!<\! 0 }$), the friction force leads to a diffusion towards smaller ${ |L_{\rt}| }$, i.e. towards larger eccentricities. This dichotomy is related to the ``secular dynamical anti-friction'' put forward in~\cite{Madigan2012}. Should one consider a quasi-Keplerian disc made of both prograde and retrograde stars, the condition ${ \partial F_{\star} / \partial L \!>\! 0 }$ would not hold anymore, and the direction of dynamical friction cannot be predicted using equation~\eqref{sign_friction}. Similarly, in ${3D}$ quasi-Keplerian systems, the associated Landau equation does not impose anymore the restriction to ${ 1 \!:\! 1 }$ resonances, which also prevents relying on equation~\eqref{sign_friction}.

\section{Conclusion}
\label{sec:conclusion}

We investigated the secular dynamics of a razor-thin axisymmetric discrete quasi-Keplerian disc surrounding a supermassive BH. In the limit where collective effects are not accounted for, such an evolution induced by finite${-N}$ effects is described by the degenerate inhomogeneous Landau equation~\eqref{Landau_Kep_disc}, recently derived in~\cite{SridharTouma2016III} and~\cite{FouvryPichonMagorrian2017}. This is the master equation of resonant relaxation~\citep{RauchTremaine1996}. The present paper presented the first effective implementation of this kinetic equation to quasi-Keplerian systems.\footnote{More generally, it has only been applied to a handful of systems: ${2D}$ razor-thin non-degenerate stellar discs~\citep{FouvryPichonChavanis2015,FouvryPichonMagorrianChavanis2015}, ${3D}$ thickened stellar discs~\citep{FouvryPichonChavanisMonk2017}, or to the ${1D}$ inhomogeneous HMF model~\citep{BenettiMarcos2017}.}

In section~\ref{sec:selfRRdisc}, we computed the self-consistent diffusion flux of the quasi-Keplerian disc and predicted the associated timescale of resonant relaxation. We also considered the simultaneous relaxation of two components of different individual mass, which leads to a mass segregation of the two components. For the specific disc considered here, we showed that heavier particles would diffuse towards smaller eccentricities and would therefore circularise, while lighter particles would diffuse towards larger eccentricities and therefore approach the central BH. More generally, all discs for which ${ \partial F_{\star} / \partial L \!>\! 0 }$ obey such trend.

In section~\ref{sec:LangevinBarrier}, we illustrated how the same formalism also describes the stochastic diffusion of individual particles, by relying on the associated Langevin equation~\eqref{Langevin_disc}. We identified the quenching of the diffusion in the central regions of the systems, a phemenon called the Schwarzschild barrier~\citep{Merritt2011}. This rewriting of the dynamics in terms of the diffusion of individual particles may be used to integrate forward in time the evolution of the system's DF as a whole. Hence the present method offers an effective alternative to direct ${N-}$body or Monte-Carlo methods, to integrate in time self-consistently the evolution of a system's DF driven by resonant relaxation. Most of the tools presented in this paper could also be implemented in the context of protoplanetary debris discs~\citep{Tremaine1998}.

The present work should be extended in various ways. It is currently limited to razor-thin axisymmetric discs for which the kinetic equation~\eqref{Landau_Kep_disc} takes a simpler form. In particular, it only involves ${ 1 \!:\! 1 }$ resonances on the precession frequencies. As shown in~\cite{FouvryPichonMagorrian2017}, ${ 3D }$ spherical systems are also quasi-stationary states whose resonant relaxation can be described by a very similar inhomogeneous degenerate kinetic equation. However, because of the additional vertical dimension, higher order resonances are allowed. For such systems, following figure~\ref{figDriftDiffLangevin}, one should investigate how the resonant diffusion quenches in the central regions and how populations of different masses may segregate in eccentricities. We accounted only for the 1PN Schwarzschild in-plane relativistic precession. It might be of interest to investigate the possible effects associated with the 1.5PN Lense-Thirring relativistic precession, which can in particular induce a precession of the wire's orbital plane. The kinetic equations considered rely on the orbit-averaging of the fast Keplerian motions and can only account for resonant diffusion. As such, it cannot capture mean motion resonances. A subsequent improvement would be to add the secondary effects of two-body non-resonant relaxation in the Langevin equation~\eqref{Langevin_disc}. In particular, this ${2-}$body non-resonant relaxation allows particles to change of energy, i.e. to change their fast action $I$, which cannot occur via resonant relaxation~\citep{EMRI}. Another venue would be to consider a central binary black hole and its orbiting stellar cluster~\citep[following e.g.][]{Rothe2010}, where the corresponding extra internal orbital degree of freedom may provide a range of intermediate frequencies, allowing the stars to pass the barrier and/or the binary to tighten, leading to e.g. EMRI. Predicting the impact of resonant relaxation with the stellar cluster to the corresponding rates should be of interest when preparing for LISA.

This paper implemented the inhomogeneous Landau equation while neglecting collective effects. In order to account for the self-gravitating amplification of the system one should rely on~\cite{FouvryPichonMagorrian2017}, which derived the corresponding inhomogeneous degenerate Balescu-Lenard equation.
For quasi-Keplerian systems, accounting for collective effects requires the evaluation of the disc's averaged response matrix, the quantity which characterises the strength of the self-gravitating amplification in the system (see~\cite{Tremaine2005,PolyachenkoShukhman2007,JalaliTremaine2012} for examples of stability investigations in the quasi-Keplerian context). Because of the BH's prevalence on the dynamics of individual stars, it is not straightforward to determine the amplitude of the gravitational polarisation in these degenerate systems, seen as a collection of Keplerian wires. This will be the subject of a future work.

\begin{acknowledgements}
We thank Scott Tremaine, Ben Bar-Or, James Binney, John Magorrian and Marta Volonteri for stimulating discussions.
Support for Program number HST-HF2-51374.001-A was provided by NASA through a grant from the Space Telescope Science Institute, which is operated by the Association of Universities for Research in Astronomy, Incorporated, under NASA contract NAS5-26555.
This research is also part of ANR grant Spin(e) (ANR-13-BS05-0005, \url{http://cosmicorigin.org}).
This work has made use of the Horizon cluster hosted by the Institut d'Astrophysique de Paris
 (we thank S.~Rouberol for running it smoothly for us), as well as of the Hyperion cluster hosted by the Institute for Advanced Study.
\end{acknowledgements}

\bibliographystyle{aa}
\bibliography{references}

\appendix

\section{The wire-wire interaction potential}
\label{sec:wirewirepotential}

Let us compute the wire-wire interaction potential $\oU_{12}$ from equation~\eqref{oU_disc}. The difficulty with such a calculation is that it requires to integrate over the fast orbital angle of each of the two wires involved. This turns out to be numerically very demanding, in particular when the two wires share the same orbital plane. Fortunately, Gauss' method~\citep{ToumaTremaine2009} allows us to perform explicitly one of these two integrals. We will not repeat the calculations presented in~\cite{ToumaTremaine2009}, but will rather detail how they may be adapted to the present context.

In order to avoid divergences associated with crossing orbits or identical orbits, the pairwise interaction potential is softened according to equation~\eqref{definition_U_softened}, for which the method of~\cite{ToumaTremaine2009} can also be applied. Using the notations from equation~\eqref{definition_U_softened}, the interaction potential from equation~\eqref{oU_disc} requires us to evaluate
\begin{equation}
\oU_{12} (\bR_{1} , \bR_{2}) \!=\! - \frac{G M_{\bullet}}{(2 \pi)^{2}} \!\!\! \int \!\!\! \rd w_{1} \rd w_{2} \, \frac{1}{\sqrt{ \big| \bm{x}_{1} [\bR_{1} , g_{1}] \!-\! \bm{x}_{2} [\bR_{2} , w_{2}] \big|^{2} \!+\! \esoft^{2}}} \, .  \nonumber
\label{start_ww}
\end{equation}
In order to emphasise the fact that one of the two angle integrals will be performed analytically, let us rewrite this equation as
\begin{equation}
\oU_{12} (\bR_{1} , \bR_{2}) = \frac{1}{2 \pi} \!\! \int \!\! \rd w_{1} \, \oUr ( \bm{x} [\bR_{1} , w_{1}] , \bR_{2} ) \, ,
\label{cal_I_ww}
\end{equation}
where ${ \oUr (\bm{x}_{1} , \bR_{2}) }$ was introduced as
\begin{equation}
\oUr (\bm{x}_{1} , \bR_{2}) = - \frac{G M_{\bullet}}{2 \pi} \!\! \int \!\! \rd w_{2} \, \frac{1}{\sqrt{\big| \bm{x}_{1} \!-\! \bm{x}_{2} [\bR_{2} , w_{2}] \big|^{2} \!+\! \esoft^{2}}} \, .
\label{definition_oUr}
\end{equation}
Here, the potential ${ \oUr (\bm{x}_{1} , \bR_{2}) }$ corresponds to the potential induced at position $\bm{x}_{1}$ by the wire of coordinates $\bR_{2}$. This potential involves an average over the orbital phase $w_{2}$ of the second particle, which is the integration that will be performed explicitly via Gauss' method. Given the mapping from equation~\eqref{definitions_a_e_eta}, equation~\eqref{definition_oUr} can be rewritten as an integral over the eccentric anomaly $\eta_{2}$. It becomes
\begin{equation}
\oUr (\bm{x}_{1} , \bR_{2}) = - \frac{G M_{\bullet}}{2 \pi} \!\! \int \!\! \rd \eta_{2} \, \frac{1 \!-\! e_{2} \cos (\eta_{2})}{\Delta} \, ,
\label{term_I_oUr}
\end{equation}
where the distance $\Delta$ is introduced as
\begin{equation}
\Delta^{2} = \big| \bm{x}_{1} \!-\! \bm{x}_{2} [\bR_{2} , \eta_{2}] \big|^{2} + \esoft^{2} \, .
\label{definition_Delta_ww}
\end{equation}
The non-trivial dependence of $\Delta$ with $\eta_{2}$ is the reason for the difficulty of computing equation~\eqref{term_I_oUr}. Let us first rewrite the distance $\Delta$ in a simpler manner. One can note that the angle-action mapping from equation~\eqref{mapping_xy} takes the form
\begin{equation}
\bm{x} [\bR , \eta] = \bm{\mathcal{R}} (g) \!\cdot\! \bm{t} (\bm{J} , \eta) \, ,
\label{mapping_rot}
\end{equation}
where the rotation matrix ${ \bm{\mathcal{R}} (g) }$ and the vector ${ \bm{t} (\bm{J} , \eta) }$ (independent of $g$) read
\begin{equation}
\bm{\mathcal{R}} (g) \!=\!
\begin{bmatrix}
\cos (g) & - \sin (g)
\\
\sin (g) & \cos (g)
\end{bmatrix}
\; ; \;
\bm{t} (\bm{J} , \eta) \!=\! 
\begin{bmatrix}
a (\cos (\eta) \!-\! e)
\\
a \sqrt{1 \!-\! e^{2}} \sin (\eta)
\end{bmatrix}
\, .
\label{definition_R_g}
\end{equation}
If the location $\bm{x}_{1}$ considered in equation~\eqref{term_I_oUr} is associated with the angle-action coordinates ${ (\bR_{1} , \eta_{1}) }$, one can then write
\begin{align}
\big| \bm{x}_{1} \!-\! \bm{x}_{2} \big| & \, = \big| \bm{\mathcal{R}} (g_{1}) \!\cdot\! \bm{t}_{1} - \bm{\mathcal{R}} (g_{2}) \!\cdot\! \bm{t}_{2} \big| = \big| \bm{\mathcal{R}} (g_{1} \!-\! g_{2}) \!\cdot\! \bm{t}_{1} - \bm{t}_{2} \big| \, .
\label{distance_rot}
\end{align}
From equation~\eqref{distance_rot}, one recovers again that the wire-wire interaction potential only depends on the phase difference ${ \Delta g \!=\! g_{1} \!-\! g_{2} }$, as in equation~\eqref{definition_A_disc}. Introducing the notation
\begin{equation}
\widetilde{\bm{x}}_{1} = \bm{\mathcal{R}} (g_{1} \!-\! g_{2}) \!\cdot\! \bm{t}_{1} = (x , y) \, ,
\label{notation_x}
\end{equation}
one can finally rewrite the distance $\Delta^{2}$ from equation~\eqref{definition_Delta_ww} as
\begin{align}
\Delta^{2} & \, = \big| \widetilde{\bm{x}}_{1} \!-\! \bm{t}_{2} \big|^{2} + \esoft^{2}  \nonumber
\\
& \, = A \!-\! 2 B \cos (\eta_{2} \!-\! \epsilon) + C \cos^{2} (\eta_{2}) \, , 
\label{calculation_Delta_ww}
\end{align}
given the quantities
\begin{align}
& \, A = x^{2} + y^{2} + a_{2}^{2} + 2 a_{2} e_{2} x + \esoft^{2} \;\; ; \;\;
 \, B \cos (\epsilon) = a_{2}^{2} e_{2} + a_{2} x \, ,  \nonumber
 \\
& B \sin (\epsilon) = a_{2} y \sqrt{1 \!-\! e_{2}^{2}} \;\; ; \;\; C = a_{2}^{2} e_{2}^{2} \, .
\label{definition_A_B_C_ww}
\end{align}
Note the presence in equation~\eqref{calculation_Delta_ww} of the quadratic term in ${ C \cos^{2} (\eta_{2}) }$. This term is the reason why one cannot apply Gauss' method to get an explicit expression for the potential $\oUr$ from equation~\eqref{term_I_oUr}. However, if instead of the potential, one considers the force by differentiating w.r.t. $\widetilde{\bm{x}}_{1}$, since $C$ is independent of $\widetilde{\bm{x}}_{1}$, this quadratic term vanishes and Gauss' method may be applied to obtain an explicit expression for the force. Equation~\eqref{term_I_oUr} gives us
\begin{align}
- \frac{\partial \oUr (\widetilde{\bm{x}}_{1} , \bR_{2})}{\partial \widetilde{\bm{x}}_{1}} = \frac{G M_{\bullet}}{2 \pi} \!\! & \, \int \!\! \rd \eta_{2} \, \frac{1 \!-\! e_{2} \cos (\eta_{2})}{\Delta^{3}}  \nonumber
\\
& \, \times \big[ \bm{F}_{0} \!+\! \bm{F}_{1} \sin(\eta_{2}) \!+\! \bm{F}_{2} \cos (\eta_{2}) \big] \, ,
\label{force_ww}
\end{align}
where the vectors $\bm{F}_{0}$, $\bm{F}_{1}$, and $\bm{F}_{2}$ obey
\begin{equation}
\bm{F}_{0} = 
\begin{bmatrix}
- x \!-\! a_{2} e_{2}
\\
- y
\end{bmatrix}
\;\; ; \;\;
\bm{F}_{1} = 
\begin{bmatrix}
0
\\
a_{2} \sqrt{1 \!-\! e_{2}^{2}}
\end{bmatrix}
\;\; ; \;\; 
\bm{F}_{2} =
\begin{bmatrix}
a_{2}
\\
0
\end{bmatrix}
\, .
\label{definition_F}
\end{equation}
Equation~\eqref{force_ww} gives the force created at the position $\widetilde{\bm{x}}_{1}$ by the wire of coordinates $\bR_{2}$. Using Gauss' method, this force may be computed analytically and is given by equation~{(67)} of~\cite{ToumaTremaine2009}, to which we refer.

Once the force from equation~\eqref{force_ww} has been computed, one may finally compute the wire-wire interaction potential from equation~\eqref{definition_oUr}. Recalling the definition of $\widetilde{\bm{x}}_{1}$ from equation~\eqref{notation_x}, the interaction potential from equation~\eqref{cal_I_ww} may be rewritten as
\begin{align}
\oU_{12} \big[ \bR_{1} , \bR_{2} \big] & \, = \frac{1}{2 \pi} \!\! \int \!\! \rd \eta_{1} \, (1 \!-\! e_{1} \cos (\eta_{1})) \, \oUr \big[ \bm{\mathcal{R}} (g_{1} \!-\! g_{2}) \!\cdot\! \bm{t}_{1} , \bR_{2} \big]  \nonumber
\\
& \!\!\!\!\!\!\!\!\!\!\!\!\!\!\!\!\!\!\!\!\!\!\!\!\! = \frac{1}{2 \pi} \bigg[ (\eta_{1} \!-\! e_{1} \sin (\eta_{1})) \, \oUr \big[ \bm{\mathcal{R}} (g_{1} \!-\! g_{2}) \!\cdot\! \bm{t}_{1} , \bR_{2} \big] \bigg]_{0}^{2 \pi}  \nonumber
\\
& \!\!\!\!\!\!\!\!\!\!\!\!\!\!\!\!\!\!\!\!\!\!\!\!\! - \frac{1}{2 \pi} \!\! \int \!\! \rd \eta_{1} \, (\eta_{1} \!-\! e_{1} \sin (\eta_{1})) \, \frac{\partial \oUr \big[ \bm{\mathcal{R}} (g_{1} \!-\! g_{2}) \!\cdot\! \bm{t}_{1} , \bR_{2} \big]}{\partial \eta_{1}}  \nonumber
\\
& \!\!\!\!\!\!\!\!\!\!\!\!\!\!\!\!\!\!\!\!\!\!\!\!\! = \oUr \big[ \bm{\mathcal{R}} (g_{1} \!-\! g_{2}) \!\cdot\! \bm{t}_{1} (\eta_{1} \!=\! 0) , \bR_{2} \big]
\label{calculation_oU_ww}
\\
& \!\!\!\!\!\!\!\!\!\!\!\!\!\!\!\!\!\!\!\!\!\!\!\!\! + \frac{1}{2 \pi} \!\! \int \!\! \rd \eta_{1} \, (\eta_{1} \!-\! e_{1} \sin (\eta_{1})) \, \frac{\partial (\bm{\mathcal{R}} (g_{1} \!-\! g_{2}) \!\cdot\! \bm{t}_{1})}{\partial \eta_{1}} \!\cdot\! \bigg[ - \frac{\partial \oUr \big[ \widetilde{\bm{x}}_{1} , \bR_{2} \big]}{\partial \widetilde{\bm{x}}_{1}} \bigg] \, ,  \nonumber
\end{align}
where the last term is given by the force from equation~\eqref{force_ww} via Gauss' method. In equation~\eqref{calculation_oU_ww}, the term involving a derivative w.r.t. $\eta_{1}$ is straightforward to compute via the mapping from equation~\eqref{definition_R_g}. The two remaining terms in equation~\eqref{calculation_oU_ww} involve both only one integration and are therefore estimated by relying on the trapezoidal rule. For a ${2\pi-}$periodic function $f$, we consider $K$ equally spaced points on ${ [0 \, ; \, 2 \pi ] }$ given by
\begin{equation}
\big[ \theta_{1} , ... , \theta_{K} \big] = \big[ 0 \, ; \, 2 \pi / K \, ; \, ... \, ; 2 \pi (K \!-\! 1) / K \big] \, .
\label{spacing_grid_trapz}
\end{equation}
Integrations are then approximated as
\begin{equation}
\!\! \int_{0}^{2 \pi} \!\!\!\! \rd \theta \, f (\theta) \simeq \frac{2 \pi}{K} \sum_{i = 1}^{K} f (\theta_{i}) \, .
\label{approx_int_trapz}
\end{equation}
In equation~\eqref{calculation_oU_ww}, the first term is estimated by sampling $K$ points in $\eta_{2}$ to compute equation~\eqref{term_I_oUr}, while the second term is estimated by sampling $K$ points in $\eta_{1}$, using the explicit expression of the force obtained from Gauss' method in equation~\eqref{force_ww}. In order to ensure an appropriate numerical convergence, the numerical applications presented in section~\ref{sec:selfRRdisc} and~\ref{sec:LangevinBarrier} used an estimation of the wire-wire interaction potential with ${ K \!=\! 10^{4} }$ sampling points.

\section{Computing the disc's surface density}
\label{sec:calcSigma}

In this Appendix, for the sake of completeness, we briefly detail how the integral from equation~\eqref{calc_init_Sigmastar} may be computed in order to determine the disc's surface density associated with a given disc's DF. To do so, we introduce the radial and tangential velocities ${ \bm{v} \!=\! (v_{\rr} , v_{\rt}) }$. The tangential velocity is given by ${ v_{\rt} \!=\! L / R }$, while the radial one satisfies
\begin{equation}
E_{\rm Kep} = \frac{1}{2} v_{\rr}^{2} + \frac{1}{2} \frac{L^{2}}{R^{2}} + \psi_{\rm Kep} (R) \, .
\label{prop_vr}
\end{equation}
Equation~\eqref{prop_vr} introduced the Keplerian potential induced by the BH as ${ \psi_{\rm Kep} (R) \!=\! - (G M_{\bullet})/R }$, while the Keplerian energy $E_{\rm Kep}$ of the particle depends only on the fast action $I$ and reads ${ E_{\rm Kep} (I) \!=\! - (1/2) (G M_{\bullet} / I)^{2} }$. One can then write
\begin{equation}
\frac{\rd v_{\rr}}{\rd I} = \frac{(G M_{\bullet})^{2}}{I^{3}} \frac{1}{\sqrt{2 (E_{\rm Kep} (I) \!-\! \psi_{\rm Kep} (R)) \!-\! L^{2} / R^{2}}} \, .
\label{dvr_dI}
\end{equation}
Paying a careful attention to the fact that the radial velocity can be both positive and negative, equation~\eqref{calc_init_Sigmastar} becomes
\begin{equation}
\Sigma_{\star} (R) \!=\! \frac{2 M_{\star} (G M_{\bullet})^{2}}{R} \!\!\! \int \!\!\! \rd L \rd I \, \frac{1}{I^{3}} \frac{F_{\star} (L , I)}{\sqrt{2 (E_{\rm Kep} (I) \!-\! \psi_{\rm Kep} (R)) \!-\! L^{2} /R^{2}}} \, .
\label{calc_SigmaStar_II}
\end{equation}
In equation~\eqref{calc_SigmaStar_II}, the integration over ${ (L,I) }$ has to be limited to the domain where the argument of the square root is positive, i.e. one must have
\begin{equation}
\frac{L^{2}}{R^{2}} + \frac{(G M_{\bullet})^{2}}{I^{2}} \leq \frac{2 G M_{\bullet}}{R} \, .
\label{constraint_Sigmastar}
\end{equation}
This first asks for the action $L$ to be such that ${ L \!\in\! [ L_{\rm min} ; L_{\rm max} ] }$, with
\begin{equation}
L_{\rm min} \!=\! 0 \;\;\; ; \;\;\; L_{\rm max} = \sqrt{2 G M_{\bullet} R} \, .
\label{constraint_L_Sigmastar}
\end{equation}
Then, for such a value of $L$, the action $I$, which also has to satisfy the constraint ${ I \!\geq\! L }$, is restricted to the domain ${ I \!\in\! [ I_{\rm min} ; I_{\rm max} ] }$, with
\begin{equation}
I_{\rm min} = \text{Max} \bigg[ L \, , \, \frac{R G M_{\bullet}}{\sqrt{2 G M_{\bullet} R \!-\! L^{2}}} \bigg] \;\;\; ; \;\;\; I_{\rm max} = + \infty \, .
\label{constraint_I_Sigmastar}
\end{equation}
Equation~\eqref{calc_SigmaStar_II} is the equation that was used in figure~\ref{figSigmastarDistorted} to compute the evolved surface density ${ \Sigma_{\star} (\tau) }$.

\balance

\end{document}